\documentclass[letterpaper,twoside,web]{latex/ieeecolor}
\usepackage[english]{babel}
\usepackage[utf8x]{inputenc}
\usepackage[T1]{fontenc}
\usepackage{bm}
\usepackage{amssymb}
\usepackage{xr}

\usepackage{amsmath}
\usepackage{empheq}
\usepackage[most]{tcolorbox}
\usepackage{graphicx}
\usepackage{epstopdf}
\usepackage{amsfonts}
\usepackage{algorithmic}
\usepackage{textcomp}
\usepackage{booktabs}

\usepackage{algorithm}
\usepackage[colorlinks=false,linkcolor=.,hypertexnames=false,hidelinks]{hyperref}

\usepackage{subcaption}

\usepackage{cite}
\usepackage{amsmath,amssymb,amsfonts}
\usepackage{algorithmic}

\usepackage{graphicx}
\usepackage{textcomp}

\usepackage{siunitx}

\DeclareMathOperator*{\argmin}{argmin}

\usepackage[export]{adjustbox}
\usepackage{latex/tmi}
\newcommand{\subparagraph}{}
\usepackage{titlesec}
\makeatletter
\def\fnum@figure{\textcolor{subsectioncolor}{\sf Fig.~\thefigure}}
\def\fnum@table{\textcolor{subsectioncolor}{\sf TABLE~\thetable}} 
\makeatother

\def\BibTeX{{\rm B\kern-.05em{\sc i\kern-.025em b}\kern-.08em
    T\kern-.1667em\lower.7ex\hbox{E}\kern-.125emX}}
\renewcommand{\theparagraph}{\arabic{subsubsection}.\arabic{paragraph}}
\titleformat{\paragraph}
{\color{nblue}\normalfont\normalsize\it}
{\color{nblue}\theparagraph}{1em}{}

\makeatletter
\renewcommand{\p@paragraph}{\thesubsection.}
\makeatother

\newcommand\copyrighttext{%
  \scriptsize \textcopyright 2021 IEEE.  Personal use of this material is permitted. Permission from IEEE must be obtained for all other uses, in any current or future media, including reprinting/republishing this material for advertising or promotional purposes, creating new collective works, for resale or redistribution to servers or lists, or reuse of any copyrighted component of this work in other works.}
  
\newcommand\copyrightnotice{%
\begin{tikzpicture}[remember picture,overlay]
\node[anchor=south,yshift=2pt] at (current page.south) {\fbox{\parbox{\dimexpr\textwidth-\fboxsep-\fboxrule\relax}{\copyrighttext}}};
\end{tikzpicture}%
}

\begin{document}

\title{Real-time non-rigid 3D respiratory motion estimation for MR-guided radiotherapy using MR-MOTUS}
\author{\small Niek R. F. Huttinga, Tom Bruijnen, Cornelis A. T. van den Berg, Alessandro Sbrizzi \thanks{The authors are with the Computational Imaging Group for MR-therapy and Diagnostics from the Department of Radiotherapy at the University Medical Center Utrecht,
Heidelberglaan 100, 3584 CX, Utrecht, The Netherlands. Correspondence should be directed to N.R.F. Huttinga (e-mail: n.r.f.huttinga@umcutrecht.nl)} \thanks{This work was supported in part by the Dutch Research Council (NWO) under Grant 15115.} \thanks{This manuscript has supplementary files which can be downloaded at \url{https://surfdrive.surf.nl/files/index.php/s/vz2xmwliglRmcjo}. The files include a document with supporting figures, and six videos that show reconstruction results. See Appendix I for more details.}}

\newcommand{\bv}[1]{\mathbf{#1}}
\newcommand{\Grad}{\mbox{\boldmath $\nabla$}}
\newcommand{\Div}{\Grad\!\cdot}
\newcommand{\Lap}{\mbox{\boldmath $\Delta$}}
\newcommand{\Curl}{\Grad\!\times}
\newcommand{\delsq}{\nabla^2}
\newcommand{\mean}[1]{\left\langle{#1}\right\rangle}
\newcommand{\infd}{\textrm{d}}
\newcommand{\diff}[2]{\frac{\infd #1}{\infd #2}}
\newcommand{\difftwo}[2]{\frac{\infd^2 {#1}}{\infd {#2}^2}}
\newcommand{\pdiff}[2]{\frac{\partial #1}{\partial #2}}
\newcommand{\pdiffcon}[3]{\left(\frac{\partial #1}{\partial #2}\right)_{#3}}
\newcommand{\pdifftwo}[2]{\frac{\partial^2 {#1}}{\partial {#2}^2}}
\newcommand{\pdifftwomix}[3]{\frac{\partial^2 {#1}}{\partial {#2}\,\partial {#3}}}
\newcommand{\vel}{\bv{v}}
\newcommand{\expu}[1]{\textrm{e}^{#1}}
\newcommand{\ehat}{\bv{e}}
\newcommand{\intomega}{ \mathop{}\!\int\limits_{\Omega}}
\newcommand{\intelement}{ \mathop{}\!\int\limits_{e_k}}
\newcommand{\intelementgen}{ \mathop{}\!\int\limits_{e_k^{xy}}}
\newcommand{\intelementref}{ \mathop{}\!\int\limits_{e_k^{\xi \eta}}}
\newcommand{\be}{\begin{equation}}
\newcommand{\ee}{\end{equation}}
\newcommand{\baa}{\begin{alignat}{2}}
\newcommand{\eaa}{\end{alignat}}
\newcommand\smathcal[1]{
  \mathchoice
    {{\scriptstyle\mathcal{#1}}}
    {{\scriptstyle\mathcal{#1}}}
    {{\scriptscriptstyle\mathcal{#1}}}
    {\scalebox{.7}{$\scriptscriptstyle\mathcal{#1}$}}
  }
\newcommand\sbmathcal[1]{
  \mathchoice
    {{\scriptstyle{\bm{\mathcal{#1}}}}}
    {{\scriptstyle{\bm{\mathcal{#1}}}}}
    {{\scriptscriptstyle{\bm{\mathcal{#1}}}}}
    {\scalebox{.7}{$\scriptscriptstyle{\bm{\mathcal{#1}}}$}}
  }

\newcommand{\note}[2]{ {\bf #1: }#2} 

\renewcommand{\sectionautorefname}{Section}
\renewcommand{\figureautorefname}{Figure}
\let\subsectionautorefname\sectionautorefname
\let\subsubsectionautorefname\sectionautorefname
\let\paragraphautorefname\sectionautorefname

\definecolor{lightgray}{rgb}{0.925,0.925,0.925}

\def\equationautorefname#1#2\null{%
  Eq.#1[#2\null]%
}

\newcommand*{\vertbar}{\rule[-1ex]{0.5pt}{2.5ex}}
\newcommand*{\horzbar}{\rule[.5ex]{2.5ex}{0.5pt}}
\newcommand{\refappendix}[1]{\hyperref[#1]{Appendix~\ref*{#1}}}
\newcommand{\refsupinf}[1]{\hyperref[#1]{Supporting Information \ref*{#1}}}
\newcommand{\refundef}{{\color{red}[x]}}

\newcommand{\beginsupplement}{%
    \setcounter{figure}{0}
    \setcounter{equation}{0}
    \renewcommand{\theequation}{S\arabic{equation}}

    \crefalias{section}{SuppInfoSection}
    \setcounter{section}{0}
}

\newcommand{\setsuppvideo}{%
    \renewcommand*{\figurename}{\hspace{-.3em}}
    \renewcommand*{\thefigure}{Supporting Information Video S\arabic{figure}}
}

\newcommand{\setsuppfig}{%
    \renewcommand*{\figurename}{\hspace{-0.3em}}
    \renewcommand*{\thefigure}{Supporting Information Figure S\arabic{figure}}
}

\newcommand{\colorred}[1]{{\color{red}{#1}}}

\renewcommand{\listfigurename}{List of figures}

\maketitle

\copyrightnotice

\begin{abstract}
The MR-Linac is a combination of an MR-scanner and radiotherapy linear accelerator (Linac) which holds the promise to increase the precision of radiotherapy treatments with MR-guided radiotherapy by monitoring motion during radiotherapy with MRI, and adjusting the radiotherapy plan accordingly. Optimal MR-guidance for respiratory motion during radiotherapy requires MR-based 3D motion estimation with a latency of 200-500 ms. Currently this is still challenging since typical methods rely on MR-images, and are therefore limited by the 3D MR-imaging latency. In this work, we present a method to perform non-rigid 3D respiratory motion estimation with 170 ms latency, including both acquisition and reconstruction. The proposed method called real-time low-rank MR-MOTUS reconstructs motion-fields directly from $k$-space data, and leverages an explicit low-rank decomposition of motion-fields to split the large scale 3D+t motion-field reconstruction problem posed in our previous work into two parts: (I) a medium-scale offline preparation phase and (II) a small-scale online inference phase which exploits the results of the offline phase for real-time computations. The method was validated on free-breathing data of five volunteers, acquired with a 1.5T Elekta Unity MR-Linac. Results show that the reconstructed 3D motion-field are anatomically plausible, highly correlated with a self-navigation motion surrogate ($\boldmath{R=0.975 \pm 0.0110}$), and can be reconstructed with a total latency of 170 ms that is sufficient for real-time MR-guided abdominal radiotherapy.
\end{abstract}


\section{Introduction}
The physiological movement of organs during radiotherapy is a source of uncertainty, and generally reduces the precision of the treatments. Such motion can typically be related to respiration, digestion and cardiac contractions. Recently, the MR-Linac was introduced as a combination of an MR-scanner and a radiotherapy linear accelerator (Linac) \cite{raaymakers2009integrating,Keall2014,Mutic2014,Lagendijk2008}. The MR-Linac holds the promise to increase the precision of radiotherapy treatments through MR-guided radiotherapy (MRgRT) by monitoring physiological motion with the MRI, and performing corresponding radiotherapy plan adaptations on the Linac. MRgRT consists of a two-fold implementation: inter-fraction and intra-fraction MRgRT. In inter-fraction MRgRT, the radiation plan should be adjusted between treatments based on the daily changes in anatomy detected with pre-treatment MRI \cite{Winkel2019}. Inter-fraction MRgRT is feasible with the currently available techniques, and is already being applied in the clinic with success \cite{Raaymakers2017,Werensteijn-Honingh2019}. In intra-fraction real-time adaptive MRgRT, MR-based motion monitoring and radiation plan adaptations should be performed during the treatments in a real-time loop, for which the required latency is determined by the speed of the targeted motion. A maximum total latency (i.e. acquisition and tracking) of 200-500 ms would be required for 3D motion estimation to achieve the full potential of the MR-Linac and optimally compensate for respiratory motion with MRgRT \cite{keall2006management,Murphy2002}. However, due to the relatively slow imaging speed of an MR-scanner, achieving this latency for 3D motion estimation is still a technical challenge.

Several respiratory motion estimation methods have been proposed in the context of MRgRT over the last years, most of which estimate motion from images. Unfortunately, these methods are directly limited by the latency of MR-imaging, including both acquisition and reconstruction time. To circumvent this, a typical approach is to estimate 3D motion from lower-dimensional cine-MR-images. For example, orthogonal 2D cine-images (i.e. 2.5D) can be acquired in order to estimate 3D motion at high temporal resolution \cite{Bjerre2013,Paganelli2018,Mickevicius2017}. In \cite{stemkens2016image}, a pre-trained 3D motion model was fit to incoming 2D cine-images to obtain fast motion estimates. Additionally, surrogate-driven motion models have been proposed that relate cine-MRI-derived surrogate signals to motion-fields \cite{Tran2020,Mcclelland2017}. Another recent work \cite{Feng2020} rapidly generates 3D MR-images by determining the best match between the current 1D motion state and 1D motion states in a 3D+t respiratory-resolved image reconstruction. For a more detailed review of related methods for MRgRT we refer to \cite{paganelli2018mri,stemkens2018nuts}, and for an in-silico comparison of several related methods we refer to \cite{paganelli2019time}.

In this work, we focus on the aforementioned technical challenge in MRgRT and propose real-time low-rank MR-MOTUS\cite{huttinga2020mr,Huttinga2021} for real-time estimation of non-rigid 3D respiratory motion-fields directly from prospectively undersampled 3D $k$-space data, with a total latency (acquisition and reconstruction) below 200 ms. The MR-MOTUS signal model relates motion-fields and a reference image to $k$-space data, allowing to reconstruct motion-fields directly from $k$-space data, given a reference image \cite{huttinga2020mr}. In \cite{Huttinga2021}, MR-MOTUS was extended to reconstruct 3D+t motion-fields in the order of minutes, using an explicit low-rank factorization in static spatial motion components and dynamic temporal motion components. Here, the same low-rank factorization is assumed, and the framework is extended to perform real-time computations by observing that a low-rank factorization allows to split the reconstruction in two phases: (I) an offline preparation phase that separates static spatial motion components from dynamic temporal motion components, and (II) an online phase in which the pre-trained static motion components are fixed and dynamic motion components are estimated with real-time acquisitions and real-time model-based reconstructions. Real-time low-rank MR-MOTUS is applied in-silico to a digital anatomical phantom, and in-vivo to 5 volunteers whose data were acquired with an MR-Linac. The reconstructed motion-fields are validated in-silico in terms of end-point-errors with ground-truth motion-fields, and in-vivo in terms of anatomical plausibility, accuracy and correspondence with a conventional respiratory-resolved compressed sensing reconstruction.

\section{Theory}

\subsection{Background MR-MOTUS}
\subsubsection{Forward signal model}
We assume a general $d$-dimensional setting, with targeted case $d=3$, and we follow the convention that bold-faced characters denote vectorizations. We define $\bv{x}_0 \mapsto \bv{x}_t$ as the mappings from coordinates $\bv{x}_0\in\mathbb{R}^d$ in a reference image to new locations $\bv{x}_t\in\mathbb{R}^d$ at time $t$. The mappings are characterized by the motion-fields $\bv{d}_t$ through $\bv{x}_t = \bv{x}_0 + \bv{d}_t(\bv{x}_0)$. This will be written in concatenated vector-form as
\begin{equation}
    \bv{X}_t = \bv{X}_0 + \bv{D}_t,   
\end{equation}
where $\bv{X}_t,\bv{X}_0,\bv{D}_t \in \mathbb{R}^{N d \times 1}$ denote the vertical concatenations over $N$ spatial points in a $d$-dimensional setup.
The MR-MOTUS forward model \cite{huttinga2020mr} explicitly relates the motion-fields $\bv{D}_t$ and a static reference image $\bv{q}_0\in\mathbb{C}^N$ to dynamic, single-channel (and possibly non-Cartesian) k-space measurements $\bv{s}_t\in\mathbb{C}^{N_k}$: 
\begin{equation}
\bv{s}_t = \bv{F}(\bv{D}_t |\bv{q}_0) + \bm{\epsilon}_t.
\label{eq:signalmodel}
\end{equation}
Here $\bm{\epsilon}_t\in\mathbb{C}^{N_k}$ is the complex noise vector and $\bv{F}:\mathbb{R}^{Nd} \mapsto \mathbb{C}^{N_k} $ is the discretization of the forward operator defined as
\begin{equation}
\label{eq:forwardmodel}
F(\bv{d}_t)[\bv{k}] = \int_\Omega q_0(\bv{x}_0) e^{-i 2 \pi \bv{k} \cdot \left[ \bv{x}_0 + \bv{d}_t(\bv{x}_0) \right]} \ \infd \bv{x}_0,
\end{equation}
where $\bv{k}\in\mathbb{R}^d$ denotes the k-space coordinate. Motion-fields can be reconstructed directly from $k$-space measurements by exploiting the availability of a fixed reference image $\bv{q}_0$, and subsequently fitting the nonlinear signal model \autoref{eq:forwardmodel} to acquired $k$-space data. We refer the reader to our previous works \cite{huttinga2020mr,Huttinga2021} for an extensive discussion on the assumptions underlying the signal model \autoref{eq:forwardmodel}. 

\begin{figure*}[t]
    \centering
    \includegraphics[width=.95\textwidth]{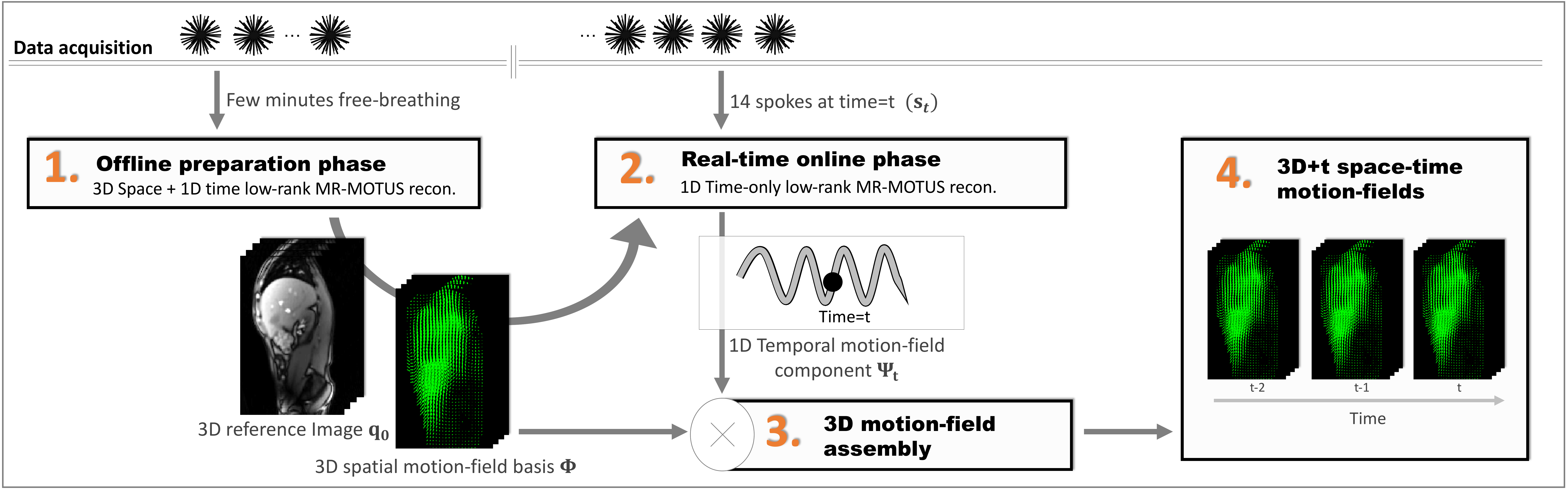}
    \caption{An overview of the real-time MR-MOTUS reconstructions, as described in \autoref{section:realtimemrmotustheory}. (1) A reference image and a spatial motion-field basis are reconstructed in an offline preparation phase from 10 minutes of data acquired in free-breathing. (2) The result of the offline phase is exploited, and only the low-dimensional dynamic representation coefficients in the spatial motion-field basis are reconstructed from just 14 spokes with a latency of just 170 ms (data acquisition + reconstruction). (3) A 3D motion-field per time instance is assembled using the spatial motion-field basis (offline) and representation coefficients (online). (4) Finally, this results in the real-time reconstruction of 3D+t motion-fields.}
    \label{fig:highleveloverview}
\end{figure*}
\subsubsection{Retrospective MR-MOTUS reconstructions of space-time motion-fields}
Reconstructing motion-fields over a longer period of time allows to exploit correlations in both space and time. However, this requires a large number of parameters, and therefore significantly increases memory consumption and reconstruction times. As shown in \cite{Huttinga2021}, a low-rank motion model can be employed that reduces memory consumption and adds a regularization in both space and time simultaneously. This model splits space-time motion-fields $\bv{D}\in\mathbb{R}^{Nd \times M}$ in a spatial component and temporal component as follows \begin{equation}\bv{D}=[\bv{D}_1,\dots,\bv{D}_M]=\bm{\Phi}\bm{\Psi}^T.\label{eq:spacetimemotionmodel}\end{equation} The first component $\bm{\Phi}$, is the spatial component that models directions and magnitude of motion per voxel. The second, $\bm{\Psi}$, is the temporal component that models the global scalings along these directions. Both components consist of $R$ rank-1 sub-components, i.e. $\bm{\Phi}\in\mathbb{R}^{Nd \times R}$ and $\bm{\Psi}=\left[\bm{\Psi}_1^T,\dots,\bm{\Psi}_M^T\right]^T\in\mathbb{R}^{M \times R}$, where $M$ denotes the number of dynamics. Since internal body motion (e.g. respiratory motion) typically occurs along similar directions over time (e.g. feet-head), this motion model allows for significantly compressed representation with $R \ll M$. Since $\text{rank}(\bm{\Phi}\bm{\Psi}_t^T)\le R \ll M$, we refer to \autoref{eq:spacetimemotionmodel} as a low-rank motion model. The motion model components can be obtained by solving the following reconstruction problem\cite{Huttinga2021}:
\begin{equation}
\label{eq:inverseproblemspatiotemporal_compressed}
\argmin_{\substack{ \bm{\Phi},\bm{\Psi}  }} \ \sum_{t=1}^M \ \left\lVert \bv{F}( \bm{\Phi}\bm{\Psi}_t^T ) - \bv{s}_t \right\rVert_2^2 +  \lambda  \mathcal{R}(\bm{\Phi}\bm{\Psi}_t^T),
\end{equation}
and can subsequently be assembled to a space-time motion-field through \autoref{eq:spacetimemotionmodel}. In \autoref{eq:inverseproblemspatiotemporal_compressed}, the first term models the data fidelity, $\mathcal{R}$ is a regularization term that incorporates a-priori assumptions, and $\lambda\in\mathbb{R}^+$ is the regularization parameter that balances both terms. The reconstruction time for \autoref{eq:inverseproblemspatiotemporal_compressed} scales with the number of dynamics $M$, which is typically large to capture large-scale dynamics. Alternatively, motion model components can be reconstructed on respiratory-resolved (rr) data,
\begin{equation}
\label{eq:inverseproblemspatiotemporal_compressed_respresolved}
\argmin_{\substack{ \bm{\Phi},\bm{\Psi}  }} \ \sum_{t=1}^{M^\textrm{rr}} \ \left\lVert \bv{F}( \bm{\Phi}\bm{\Psi}_t^T ) - \bv{s}^{\textrm{rr}}_t \right\rVert_2^2 +  \lambda  \mathcal{R}(\bm{\Phi}\bm{\Psi}_t^T),
\end{equation}
where $\bv{s}^{\textrm{rr}}_t$ denotes $k$-space data that is sorted into $M^\textrm{rr}$ respiratory phases. Since typically $M^\textrm{rr} \ll M$, and the number of readouts per dynamic in $|\bv{s}^{\textrm{rr}}_t|$ is larger than in $|\bv{s}_t|$, this results in a reconstruction problem with better conditioning and a reduced reconstruction time that is beneficial in practice. Due to the practical benefits we will consider \autoref{eq:inverseproblemspatiotemporal_compressed_respresolved} in this work.

\subsection{Extension to real-time reconstructions: framework overview}
\label{section:realtimemrmotustheory}
In \cite{Huttinga2021} it was shown that high temporal resolution space-time motion-fields can be reconstructed by solving \autoref{eq:inverseproblemspatiotemporal_compressed} or \autoref{eq:inverseproblemspatiotemporal_compressed_respresolved}. However, despite the low-rank factorization and respiratory sorting,  \autoref{eq:inverseproblemspatiotemporal_compressed_respresolved} is still a medium-scale reconstruction problem, resulting in reconstruction times in the order of minutes. Retrospectively, the reconstructed space-time motion-fields are valuable for e.g. the assessment of dose accumulated during treatments, but the long reconstruction times prevent the direct application of the framework to real-time MRgRT.

We observe that reconstructions with the low-rank model yield a convenient representation that allows for significant reduction in computation times. All motion-fields $\bv{D}_1,\dots\bv{D}_M$ are represented as a linear combination of the $R$ columns of $\bm{\Phi}$, with the $R$ representation coefficients given as the columns of $\bm{\Psi}$ (\autoref{eq:spacetimemotionmodel}). It has empirically been shown that realistic respiratory motion can be represented with $R=1\dots 3$, i.e. with few basis functions and few representation coefficients \cite{stemkens2016image,Huttinga2021}. 

These observations suggest a strategy to reduce the computational burden of \autoref{eq:inverseproblemspatiotemporal_compressed_respresolved} by splitting the reconstruction in two phases: I) a medium-scale offline preparation phase that reconstructs $\bm{\Phi}$ from data acquired during representative motion, and II) a small-scale inference phase that exploits the availability of $\bm{\Phi}$ and only reconstructs the few representation coefficients per dynamic in real-time:
\begin{align}
\{\bm{\Phi}^{\textrm{rr}},\bm{\Psi}^{\textrm{rr}}\}&=\argmin_{\substack{\bm{\Phi},\bm{\Psi}}} \sum_{t=1}^{M^\text{rr}} \left\lVert \bv{F}( \bm{\Phi}\bm{\Psi}_t^T ) - \bv{s}^\text{rr}_t \right\rVert_2^2 +  \lambda \textrm{TV}(\bm{\Phi}\bm{\Psi}_t^T), \label{eq:MRMOTUSoffline}\\[.2cm]
\{\bm{\Psi}_t\}&=\argmin_{\substack{\bm{\Psi}_t }} \     \lVert \bv{F}\left(\bm{\Phi}^{\textrm{rr}}\bm{\Psi}_t^T\right) - \bv{s}_t\rVert_2^2 + \mu \lVert \bm{\Psi}_t -  \bm{\Psi}_{t-1} \rVert_2^2. \label{eq:MRMOTUSonline}
\end{align}
Here $\mu>0$ is a regularization parameter that stabilizes the real-time reconstructions by penalizing large deviations from the solution at the previous dynamic. Moreover, "TV" is defined as the vectorial total variation, computed as the $L^2$-norm over the total variation per motion-field direction \cite{Blomgren1998}:
$$\textrm{TV}(\bv{D}_t) := \sqrt{\sum_{p=1\dots d} \left(\sum_i \lVert [\nabla \bv{D}_t^p]_i \rVert_2  \right)^2},  $$
where $[\nabla \ \cdot \ ]_i$ denotes the gradient at the $i$-th spatial coordinate, and the superscript $p$ denotes the motion-field direction. The first phase of the reconstruction, \autoref{eq:MRMOTUSoffline}, consists of a medium-scale reconstruction problem which can be solved offline in the order of minutes on a desktop PC, for respiratory-sorted data $\bv{s}^\text{rr}$ in a training set. The second phase, \autoref{eq:MRMOTUSonline}, consists of an extremely small-scale reconstruction problem with typically only 1-3 unknowns per dynamic, which can be solved online in the order of few milliseconds on a desktop PC, for dynamics not present in the training set. In practice, the first phase can be performed offline during the radiotherapy treatment preparation, and the second phase online with minimal latency during irradiation. \autoref{fig:highleveloverview} schematically illustrates the main steps of the workflow.

\section{Methods}
\begin{figure*}[t]
    \centering
    \includegraphics[width=.95\textwidth]{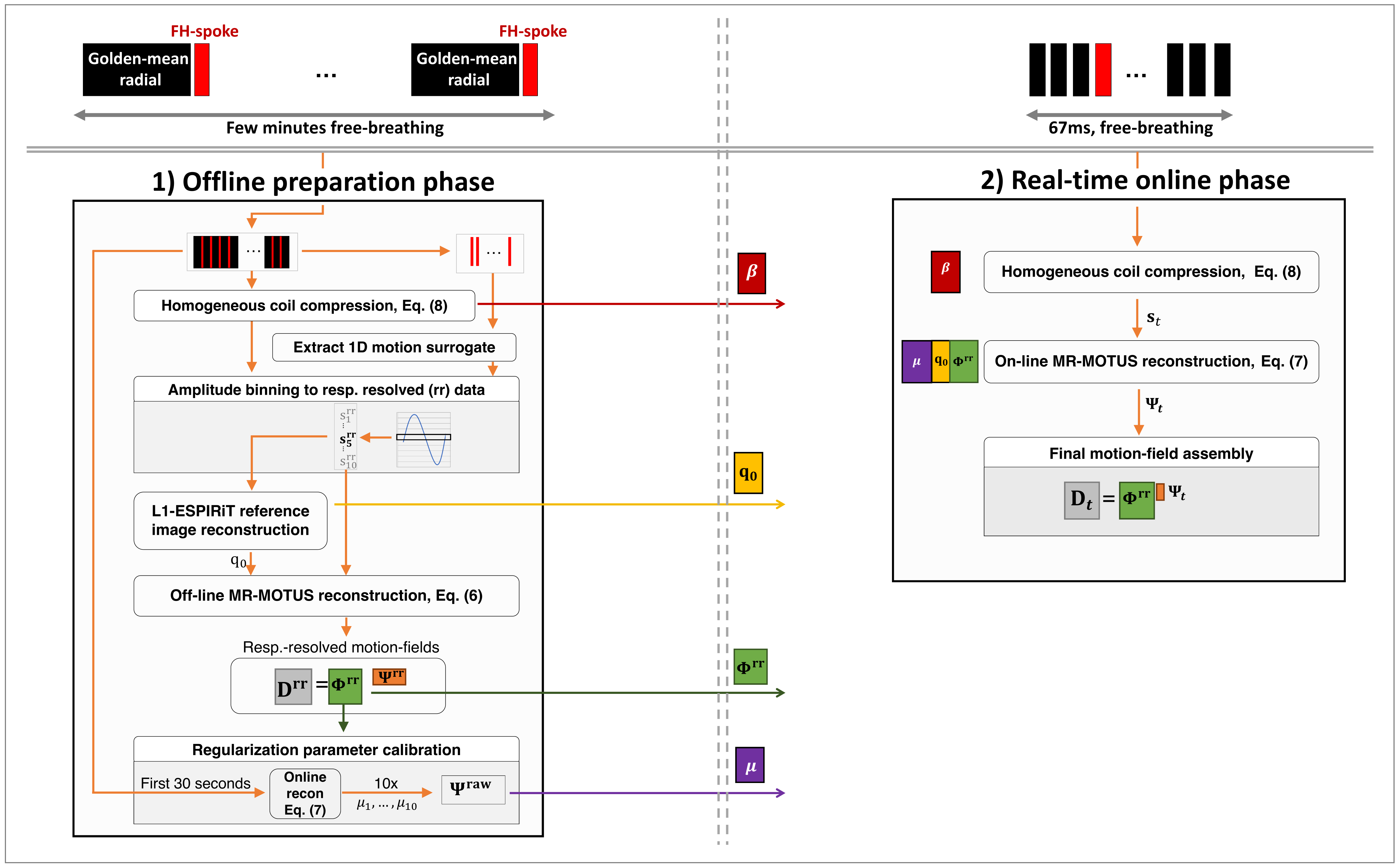}
    \caption{A detailed overview of the reconstruction pipeline. Several components are reconstructed in the offline phase and exploited in the real-time phase to reduce the computation time to 200 ms per dynamic: coil compression coefficients $\bm{\beta}$, reference image $\bv{q}$ and spatial motion-field basis $\bm{\Phi}^\textrm{rr}$. All steps in the figure are elucidated in \autoref{section:invivorealtimemethods}. }
    \label{fig:lowleveloverview}
\end{figure*}

To evaluate the (source of) local motion-field errors in the proposed method, several validation experiments were performed with a digital anatomical XCAT phantom \cite{segars20104d} with realistic respiratory motion. Real-time MR-MOTUS reconstructions were performed for 5 volunteers whose data were acquired on an MR-Linac, during free-breathing, and with a multi-channel radiolucent receive array. The anatomical plausibility of the reconstructed in-vivo motion-fields was evaluated by means of the Jacobian determinant, allowing to detect possibly unrealistic compression or expansion induced by the reconstructed motion-fields. Additionally, the global accuracy of the in-vivo motion-fields was assessed by means of the Pearson correlation and Bland-Altman difference plots between the reconstructed motion in feet-head direction, and a 1D respiratory motion surrogate. This allowed for a validation of the 3D MR-MOTUS motion-fields at high temporal resolution (6.7 Hz). Finally, the reconstructions in the offline phase were qualitatively compared to respiratory-resolved compressed sensing reconstructions in 3D. More details on the in-silico and in-vivo experiments will be discussed in, respectively, \autoref{section:insilicorealtimemethods} and \autoref{sec:realtimeinvivo_new}. We first describe the complete reconstruction pipeline in \autoref{section:invivorealtimemethods}, which is also visualized in detail in \autoref{fig:lowleveloverview}. All computations in this work were performed in Matlab 2019a (The MathWorks Inc., Naticks, Massachusetts). Representative code to perform similar reconstructions will be made available at \url{https://github.com/nrfhuttinga/Realtime_MRMOTUS}.

\subsection{Real-time reconstruction pipeline}
\label{section:invivorealtimemethods}

\subsubsection{Data acquisition/simulation}
In practice all data were acquired or simulated during free-breathing. In-vivo, a multi-element receive array was used and a steady-state spoiled gradient echo sequence (SPGR) was employed, TR = 4.8 ms, TE = 1.8 ms, FA = 20\textdegree, FOV = 30 cm $\times$ 30 cm $\times$ 30 cm, and  BW = 540 Hz. A 3D golden-mean (GM) radial kooshball trajectory \cite{chan2009temporal} was employed, interleaved every 31 spokes with a self-navigation spoke oriented along the feet-head (FH) direction. The 3D GM kooshball trajectory efficiently acquires 3D $k$-space data with relatively uniform angular distribution at all temporal resolutions \cite{chan2009temporal}, while the self-navigation spoke yields a motion surrogate every $31\cdot\textrm{TR}=148.8$ ms, allowing for retrospective validation of the reconstructed motion-fields in FH direction at 6.7 Hz temporal resolution \cite{feng2016xd,Pang2014}.

\subsubsection{Offline preparation phase}\label{section:preprocess_and_refimage}
The offline preparation phase consists of several steps, outlined below. 

\paragraph{Homogeneous coil compression}
The MR-MOTUS signal model requires single-channel $k$-space data with approximately homogeneous coil sensitivity \cite{Huttinga2021}. To achieve this, we followed \cite{Huttinga2021} and linearly compressed all data to a single virtual channel prior to MR-MOTUS reconstructions. Compression coefficients $\bm{\beta}\in\mathbb{C}^{N_c}$ were obtained by solving:  
\begin{equation}\label{eq:homogeneous_coil_compression}\min_{\boldsymbol{\beta}}\lVert{\mathbf{S}\boldsymbol{\beta}-\mathbf{1}}\rVert_2^2\quad\Rightarrow\quad{}\boldsymbol{\beta}=(\mathbf{S}^H\mathbf{S})^{-1}\mathbf{S}^H\mathbf{1}.\end{equation}
Here $N_c$ denotes the number of channels, $\bv{S}\in\mathbb{C}^{N \times N_c}$ the coil sensitivities, and $\bv{1}\in\mathbb{R}^{N}$ an all-one vector.

\paragraph{Surrogate signal extraction and binning}
A surrogate was extracted from the self-navigation spoke along FH direction for the purpose of respiratory binning and validation at high temporal resolution. For this we follow the principal component analysis (PCA) approach of \cite{feng2016xd}, and extract the surrogate as the principal component with the highest spectral density in the respiratory motion frequency range 0.1 - 0.5 Hz. A cluster of coil elements was determined for which the extracted respiratory motion surrogate signals show high correlation by following the method in \cite{Zhang2016}. The final surrogate was extracted from $k$-space data averaged over this cluster. Finally, a low-pass filter was applied to remove remaining high-frequency oscillations.

Subsequently, all acquired data were sorted based on the amplitude of the extracted motion surrogate. This required a surrogate signal value per readout, so a nearest-neighbor interpolation was performed to interpolate the motion surrogate from the temporal resolution of the self-navigation spokes to the temporal resolution of a single spoke. A total of 10 respiratory bins were selected, and we denote the resulting respiratory-resolved data as $\bv{s}^\textrm{rr}_1,\dots,\bv{s}^\textrm{rr}_{10}$.

\paragraph{Reference image reconstruction}
\label{sec:referenceimagereconstruction}
The bin in the respiratory phase halfway between exhale and inhale, i.e. mid-ventilation, containing around 12000 spokes was selected for reference image reconstruction. Two mid-ventilation reference images were reconstructed with $L^1$-wavelet ESPIRiT using the BART MRI reconstruction toolbox \cite{uecker2015berkeley} (regularization parameter: 5e-3, iterations: 550). The first is a low-resolution reference image (6.7 mm isotropic) for subsequent MR-MOTUS motion-field reconstructions, and the second a higher resolution reference image (3 mm isotropic) for visualization purposes. 

\paragraph{Offline MR-MOTUS reconstructions}
\label{paragraph:offlinemrmotusmethods}
The sorted data was used to perform a respiratory-resolved MR-MOTUS reconstruction with \autoref{eq:MRMOTUSoffline}, following \cite{Huttinga2021} and using the code that was made available online at \url{https://github.com/nrfhuttinga/LowRank_MRMOTUS}. The number of respiratory phases was set to 10, the regularization parameter $\lambda$ is empirically tuned once and then fixed for all volunteers. In accordance with previous work \cite{stemkens2016image}, the number of ranks was set to 1 ($R=1$). The motion-fields were parameterized with cubic B-spline bases in space (24 splines in AP, LR, and 16 splines in FH) and time (5 splines). All spline coefficients were randomly initialized in $[-0.5,0.5]$, and scaled such that the sets of spatial and temporal coefficients both had unit norm. The reconstruction was performed with 60 iterations of L-BFGS \cite{liu1989limited} (using the Matlab wrapper \cite{beckerlbfgsb2019}), with reconstruction times in the order of minutes. This reconstruction resulted in the spatial component $\bm{\Phi}^{\textrm{rr}}$ required for the online reconstructions. 

\subsubsection{Online inference phase}
\label{section:onlineinferencephase}
In the online inference phase, $k$-space data was grouped into dynamics, with 14 spokes per dynamic, and \autoref{eq:MRMOTUSonline} was solved per dynamic. The reconstruction for time $t$ was initialized with the reconstruction at time $t-1$, and the reconstruction problem was solved with a single iteration of a GPU-accelerated Gauss-Newton scheme. Finally, a 3D motion-field $\bv{D}_t$ was assembled per dynamic using the offline reconstructed spatial components $\bm{\Phi}^\textrm{rr}$ and the online reconstructed temporal component: $\bv{D}_t = \bm{\Phi}^\textrm{rr} \bm{\Psi}_t^T$.


Several aspects of the reconstruction were considered to speed up computations. The forward model was implemented with an explicit matrix-vector multiplication, rather than with the type-3 non-uniform FFT (NUFFT) that was used for \autoref{eq:MRMOTUSoffline} due to the relatively large overhead of the NUFFT for few $k$-space samples. The whole online phase was performed on a GPU (Nvidia Quadro K620 2GB) using MATLAB's \texttt{gpuArrays}, which resulted in a factor 8 reduction in computation time in comparison with the CPU implementation. The pseudo-code of the reconstruction algorithm and more details on the speed-up steps are provided in the Supporting Information.


The total processing time of all steps above depends on the number of spokes per dynamic. Ideally, a large number of spokes should be selected per dynamic to improve the conditioning of the reconstruction problem, but this increases both the acquisition and the reconstruction time. Therefore, a trade-off has to be made to satisfy the latency requirement of 200 ms for real-time MRgRT. The dependency of the total processing time on the number of spokes per dynamic was analyzed by calculating the mean and standard deviations over 2000 online reconstructions (see \autoref{section:invivorealtimeresults}, Supporting Information Figure S1). Based on this analysis, only 14 spokes and 8 samples per spoke were selected per dynamic, which resulted in a mean real-time reconstruction time of 103 ms per dynamic. With an acquisition time of $14\cdot\textrm{TR}=67 \si{ms}$, this resulted in a total latency of 170 $\si{ms}$, which is well below the latency requirement for real-time MRgRT \cite{Murphy2002,keall2006management}. 
\subsection{In-silico validation: error analysis of real-time 3D MR-MOTUS reconstructions}
\label{section:insilicorealtimemethods}

\begin{figure}[tbp]
    \centering
    \includegraphics[width=.48\textwidth]{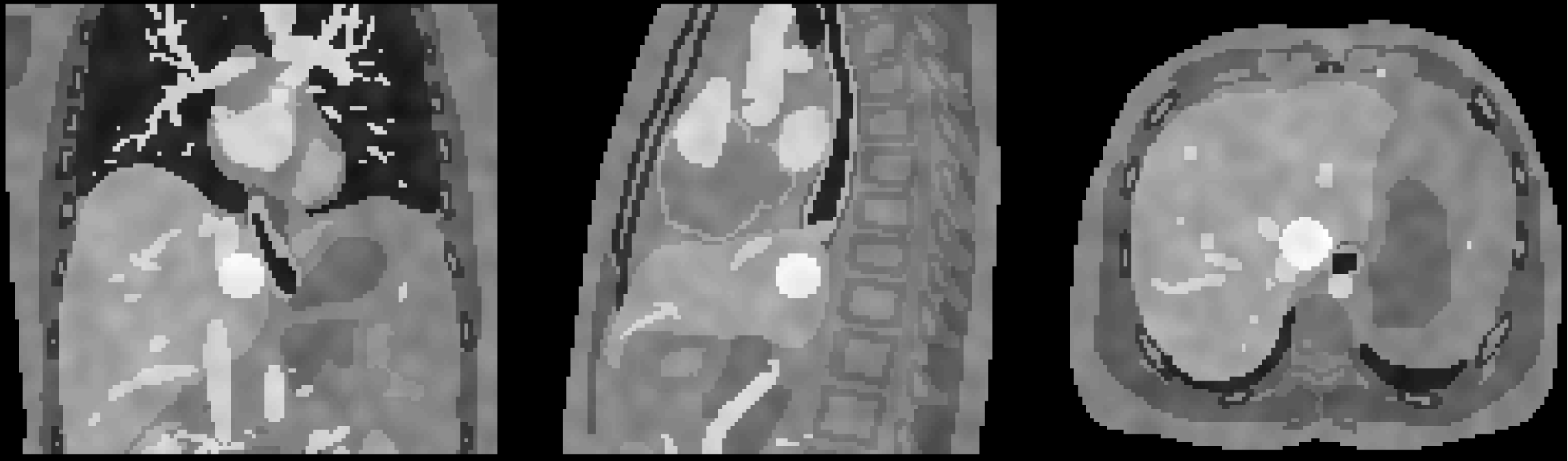}
    \caption{XCAT reference volume for the simulations described in \autoref{section:insilicorealtimemethods} (left to right: coronal, sagittal, axial) with manually added MR-contrast and a spherical lesion in the liver.}
    \label{fig:xcat_volumes}
\end{figure}

In-silico validations were performed to evaluate the local errors of the proposed two-step motion-field reconstruction approach, i.e. inferring time-resolved motion-fields using a motion model built on respiratory-resolved data. We analyzed the contributions of both the offline phase and the online phase to this error. Data was simulated using the XCAT digital phantom for respiratory motion \cite{segars20104d}, to which MR-contrast was manually added, see \autoref{fig:xcat_volumes} for the resulting XCAT phantom.

\subsubsection*{Realistic motion-fields}
The following aspects were considered to obtain realistic motion-fields. To ensure motion-fields that are invertible and consistent with the deformed XCAT images, the original XCAT motion-fields were post-processed with the recently published framework by Eiben et al \cite{Eiben2020}. To simulate lower velocity in exhale than inhale, $\cos^4$ waveforms were used as input to the XCAT framework. Hysteresis was simulated by a phase-delay between chest and abdominal input waveforms. To simulate pseudo-periodic motion, end-exhale and end-inhale position deviations were randomly generated within a range of respectively 1\% and 2\% of the waveform amplitude. No cardiac motion was applied. See \autoref{fig:xcat_waveforms} for the input waveforms. 
\subsubsection*{Translational error of the two-step reconstruction approach}
To evaluate the translational performance of the framework in case of normal breathing in the offline phase and different breathing patterns in the online phase, the offline phase was performed on respiratory-resolved data simulated during normal breathing, and the online real-time phase on data simulated during four breathing patterns: normal breathing, chest-only, abdominal-only and amplitude drifts (see \autoref{fig:xcat_waveforms}). Errors in this experiment specifically due to the offline phase were assessed by comparing real-time reconstructions with an offline reconstructed spatial basis ${\bm{\Phi}}$, and with a ground-truth ${\bm{\Phi}}$, both obtained during normal breathing. The columns of the ground-truth $\bm{\Phi}$ were obtained as the left-singular vectors of the ground-truth motion-fields during normal-breathing. 
\subsubsection*{Data simulation}
To simulate data on dynamic XCAT images we proceeded as follows. An end-exhale XCAT image was taken as the reference image, and smooth magnitude and phase variations were added to obtain a complex image with some intra-organ. This reference image was deformed with the post-processed motion-field by cubic interpolation. Each breathing phase consisted of 100 dynamics and five breathing cycles, each with a period of 5 seconds. Data was generated from the deformed images with a type-2 NUFFT \cite{barnett2018parallel} evaluated on the same trajectory as for the in-vivo experiments, i.e. a golden-mean 3D radial trajectory interleaved with a self-navigation FH spoke every 31 spokes. Complex noise was added to achieve an SNR of approximately 50, and 400 spokes were simulated per dynamic. With the current in-vivo acquisition parameters (see \autoref{sec:realtimeinvivo_new}) this equals a free-breathing acquisition time of around 3 minutes per breathing phase.

\subsubsection*{Data processing and reconstruction}
The data processing for the offline phase was kept similar to that for in-vivo reconstructions as described in \autoref{section:preprocess_and_refimage}, including the reconstruction of the reference image from binned data. Some exceptions are that the coil compression was not required since single-channel data was generated. Moreover, the reference image was reconstructed in end-exhale, and the binning and data-sorting was performed for inhale and exhale separately to increase the sensitivity for the different breathing types. For the online reconstructions, 14 spokes were simulated per dynamic, similar to the in-vivo experiments, and online reconstructions were performed as described in \autoref{section:onlineinferencephase}. A similar spline basis as for the in-vivo experiments was used, and no regularizations were employed ($\mu,\lambda=0$).

\subsubsection*{Performance evaluation}
\begin{figure*}[h]
    \centering
    \includegraphics[width=.90\textwidth,trim={1cm 0 0 -1cm}]{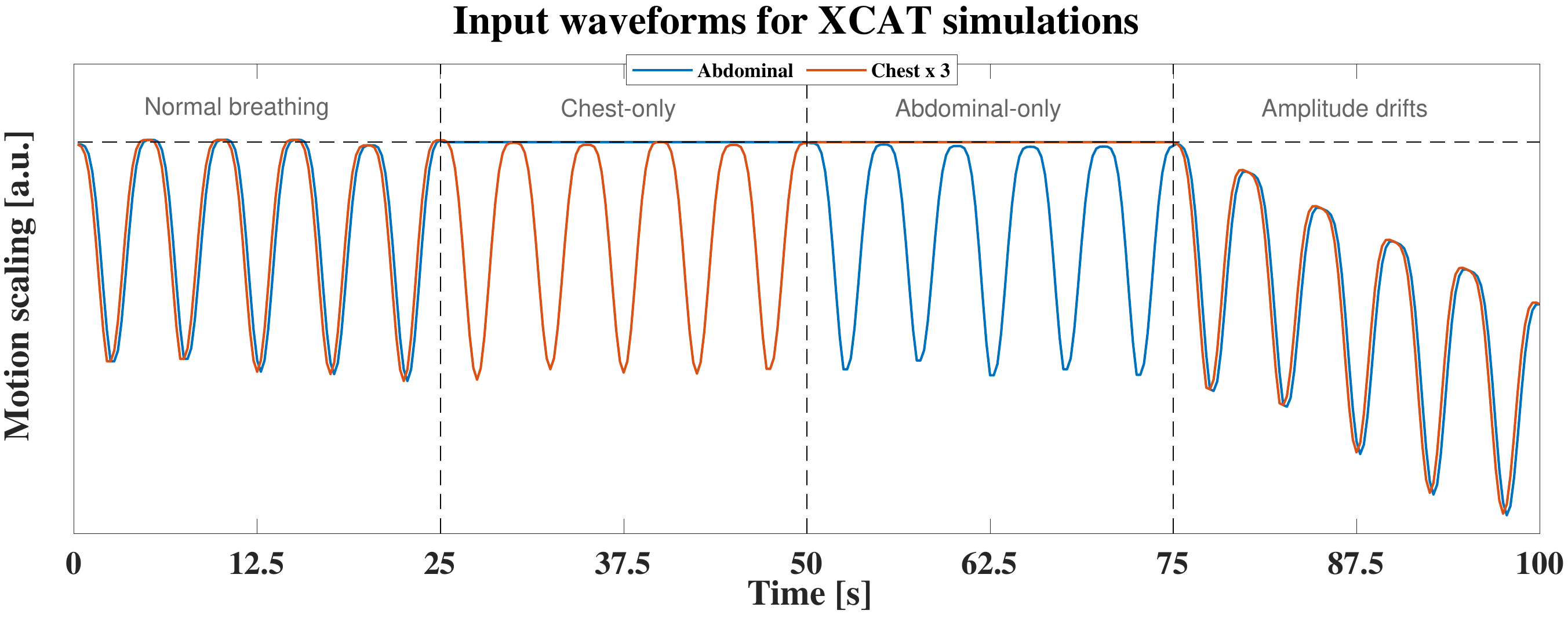}
    \caption{Input abdominal (blue) and chest (orange) waveforms for the XCAT simulation study described in \autoref{section:insilicorealtimemethods}. Here, `Chest x 3' denotes that the amplitude of the actual chest waveform is three times lower than that of the abdominal waveform, resulting in relatively smaller chest motion. Note the varying end-inhale and end-exhale positions, the phase delay causing hysteresis, and the different breathing patterns.}
    \label{fig:xcat_waveforms}
\end{figure*}

Performance in all experiments above was analyzed in terms of end-point-errors (EPEs) between reconstructed and ground-truth motion-fields, evaluating both the complete volumetric spatial distribution and the mean over an ROI defined as a spherical tumor insert in the liver. The maximum displacement of this spherical tumor insert in the normal breathing scenario in \autoref{fig:xcat_waveforms} was 14.8 mm (7 mm in AP, 13 mm in FH, 0 mm in LR), for chest-only it was 7 mm (7 mm in AP, 0 mm in FH, 0 mm in LR), and for abdominal-only 13 mm (0 mm in AP, 13 mm in FH, 0 mm in LR). For the amplitude drift scenario all displacements of the normal breathing scenario were scaled with a factor that linearly increased from 1 to 1.5, resulting in a maximum displacement of 22.1 mm (10.5 mm in AP, 19.5 mm in FH, 0 mm in LR).

\subsection{Real-time in-vivo reconstructions}
\label{sec:realtimeinvivo_new}
For the real-time in-vivo reconstruction we follow the pipeline outlined in \autoref{section:invivorealtimemethods}. Prior to these reconstructions data is acquired and processed according to the steps below.
\subsubsection*{Data acquisition}
    All data were acquired on a 1.5T MR-Linac (Elekta Unity, Elekta AB, Stockholm, Sweden) from 5 healthy volunteers (BMI $\in[19.5;27.2]$, Age $\in[22;44]$) during 11:40 minutes of free-breathing, no breathing coaching was performed. The first 10 minutes of data were used for the offline phase (\autoref{section:preprocess_and_refimage}), the last 1:40 minutes were used for the real-time phase (\autoref{section:onlineinferencephase}). All experiments were approved by the institutional review board, carried out in accordance with the relevant guidelines and regulations, and written informed consent was obtained from all volunteers prior to the experiments. 
\subsubsection*{Online reconstruction}

A total of 100 seconds of dynamics were reconstructed per volunteer, of which the first 30 seconds were used to empirically tune the regularization parameter $\mu$ (\autoref{eq:MRMOTUSonline}) per volunteer so as to remove high-frequency oscillations, without over-smoothing. 

\subsection{\color{subsectioncolor}In-vivo anatomical plausibility test with Jacobian determinants}
\label{section:jacdet_validation_methods}
Organs such as the liver, spleen and kidney consist of liquid-filled tissue structures, and can therefore be assumed incompressible and thus volume-preserving during deformation \cite{zachiu2018anatomically}. Volume change due to deformation-fields can be quantified by the determinant of the local deformation-field's Jacobian matrix, which will be referred to as the Jacobian determinant. Hence, for anatomically plausible motion-fields, the Jacobian determinant should be close to unity within organs such as the liver. 

The anatomical plausibility of the reconstructed motion-fields was therefore evaluated by means of the Jacobian determinant for the two most extreme positions with respect to the mid-ventilation reference image: end-exhale and end-inhale. The spatial derivatives required for the computation of the Jacobian determinants were obtained through finite differences.

\subsection{In-vivo global accuracy test at high temporal resolution}
\label{section:fh_validation_methods}
Alongside the anatomical plausibility, the accuracy of motion-fields is also of great importance for applications such as real-time adaptive MRgRT. However, it is evidently not straightforward to validate 3D motion-fields at high temporal resolution using MR-images, due to the relatively slow imaging speed of MRI. We propose an alternative strategy for to check the global accuracy of the motion-fields. The magnitude of respiratory motion is typically dominant in FH direction \cite{Heerkens2014,von20074d}, and its temporal profile can be extracted at high temporal resolution with PCA on self-navigation spokes along the FH direction\cite{feng2016xd,Pang2014}. The reconstructed motion-fields should resemble the same temporal profile, since the FH motion that can be extracted from self-navigation spokes along FH should also be present in the reconstructed motion-fields. For these reasons, we validated the motion-fields by correlation analyses between the temporal component of the reconstructed motion-fields $\bm{\Psi}$, and a 1D motion surrogate extracted from the self-navigation spokes along FH direction. The surrogate was extracted as described in \autoref{section:preprocess_and_refimage}. In addition to the surrogate, a 1D FFT on the FH self-navigation spokes also yielded high temporal resolution projections of the moving anatomy on the FH-axis. 

Qualitative analysis was performed by visual comparison between these projection, the 1D FH motion surrogate extracted with PCA, and the reconstructed temporal MR-MOTUS profile $\bm{\Psi}$. Since both signals represent a global scaling, they cannot be compared directly. Therefore, the signals were normalized by substracting the mean and dividing by their respective standard deviations. Additionally, Bland-Altman plots were generated to assess the similarity. Finally, quantitative analysis was performed by computing the Pearson correlation coefficient between the MR-MOTUS temporal profile and the 1D motion surrogate for all volunteers. 

\subsection{Comparison with respiratory-resolved compressed sensing}
The offline-reconstructed respiratory-resolved MR-MOTUS reconstructions were compared to offline respiratory-resolved compressed sensing reconstructions as follows. The reconstructed motion-fields were used to warp the reference image, resulting in respiratory-resolved volumetric images. A compressed sensing reconstruction was performed with BART \cite{uecker2015berkeley} on the same single-channel respiratory-binned data as was used for the offline MR-MOTUS reconstructions, also resulting in respiratory-resolved volumetric images. The reconstruction was performed over all respiratory bins simultaneously, using 550 iterations, L1-wavelet spatial, and total-variation temporal regularization. Finally, to evaluate the differences between the two reconstructions, the two sets of volumetric images were visually compared side-by-side and in terms of absolute differences.

\section{Results}
\subsection{In-silico validation: error analysis of real-time 3D MR-MOTUS reconstructions}
\label{section:insilicorealtimeresults}
\begin{figure}[tb]
    \centering
    \includegraphics[width=0.48\textwidth]{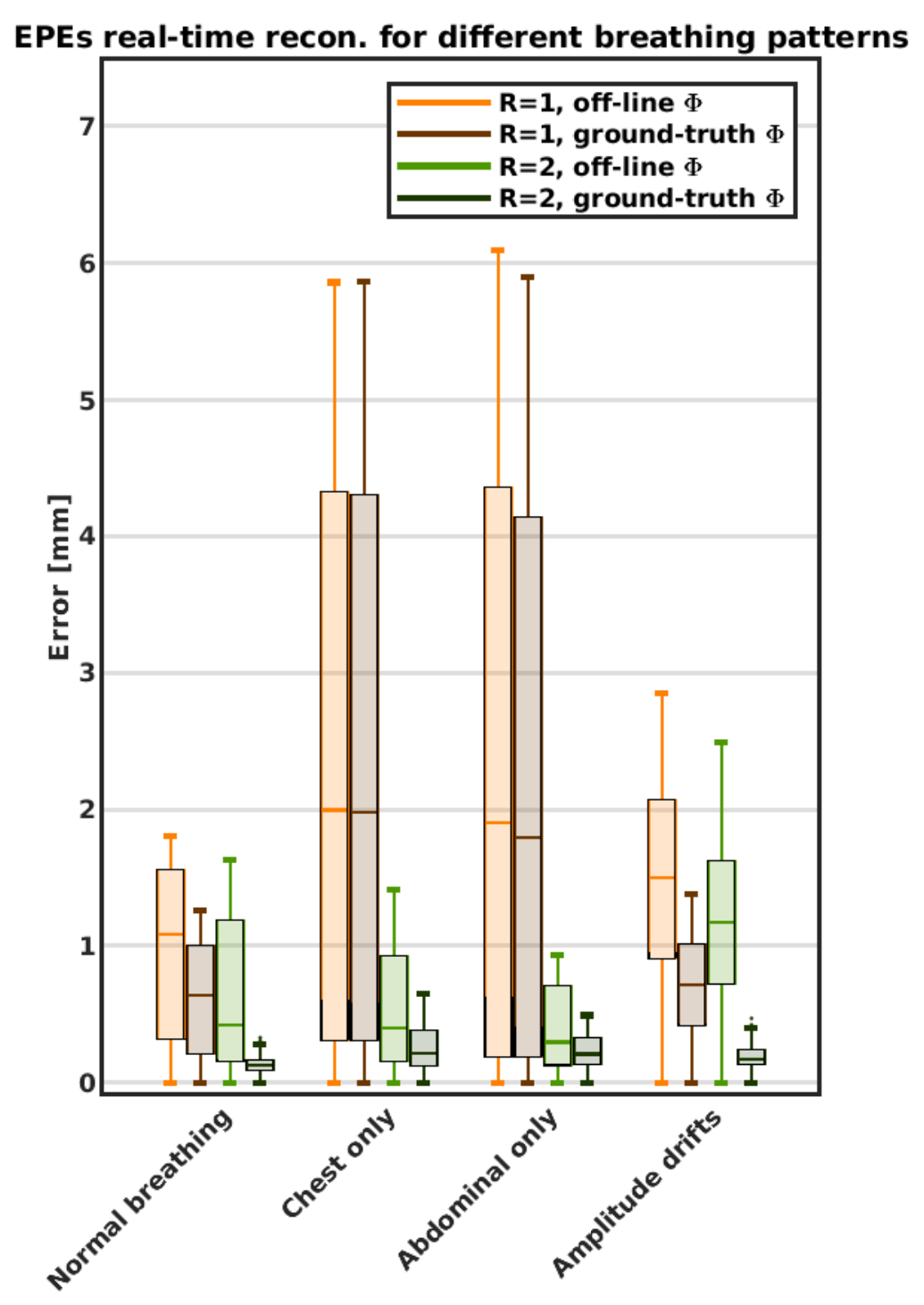}
    \caption{Mean end-point-errors (EPEs) over an ROI (tumor insert) for real-time MR-MOTUS reconstructions on simulated XCAT data (see \autoref{section:insilicorealtimemethods}, \autoref{section:insilicorealtimeresults}).}
    \label{fig:epe_roi_xcat}
\end{figure}

\begin{figure}[t]
    \centering
    \includegraphics[width=0.48\textwidth,frame,trim={-1cm 0 -1cm 0}]{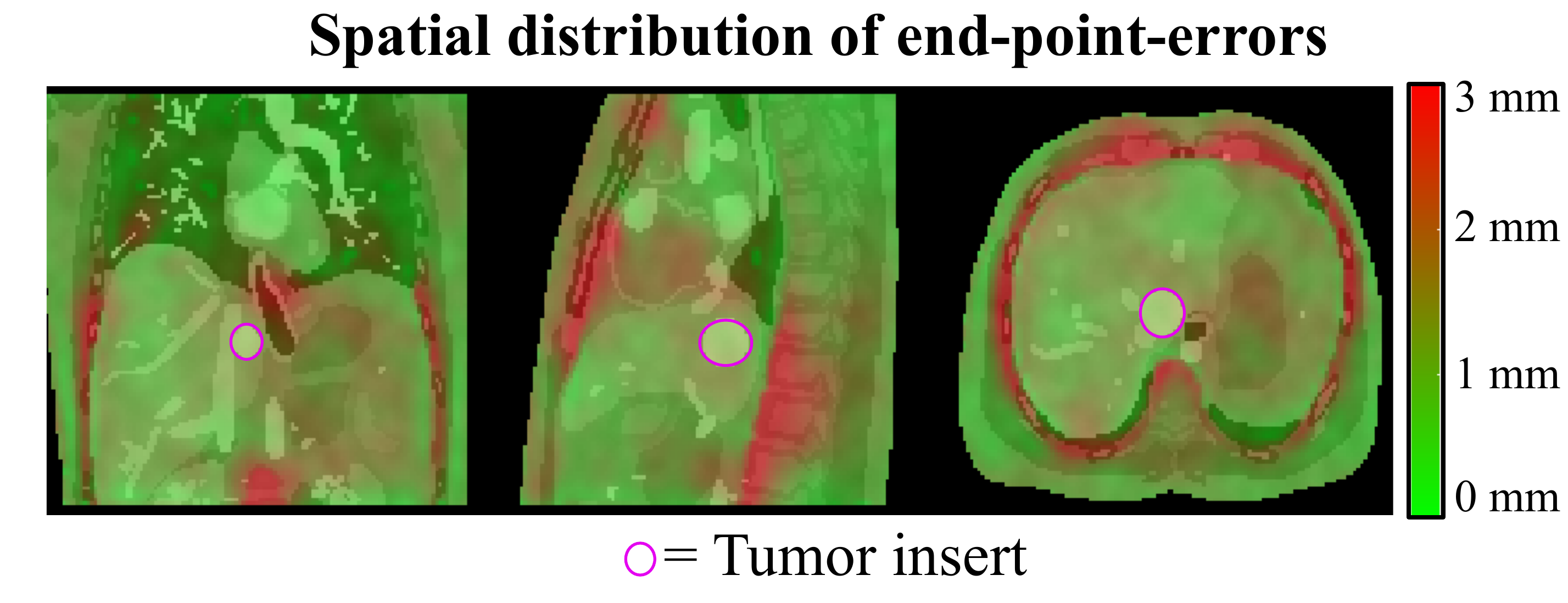}
    \caption{Spatial distribution of EPEs of real-time MR-MOTUS reconstruction with $R=2$ and offline reconstructed $\bm{\Phi}$ (see \autoref{section:insilicorealtimemethods}, \autoref{section:insilicorealtimeresults}). The dynamic with the largest errors during normal breathing was selected for visualization.}
    \label{fig:epe_spatial_distribution}
\end{figure}

The end-point-errors between the real-time MR-MOTUS XCAT reconstructions described in \autoref{section:insilicorealtimemethods} are evaluated over an ROI (spherical lesion in the liver) in \autoref{fig:epe_roi_xcat}, and over the whole volume in \autoref{fig:epe_spatial_distribution}. 

Several conclusions can be drawn from \autoref{fig:epe_roi_xcat}. The translational performance can be assessed by comparing the same colors over the different scenarios. This shows that an extreme change of breathing patterns, i.e. from normal breathing to either chest-only or abdominal-only, increases the errors in most cases. An exception is the rank-2 model with offline-reconstructed $\bm{\Phi}$, which shows an improved performance for the extreme scenarios. Comparing the same tints of the different colors per breathing pattern shows an improved performance of the rank-2 models over the rank-1 models in all cases, both with offline reconstructed and ground-truth $\bm{\Phi}$.

The performance of the complete reconstruction pipeline for the $R=1$ and $R=2$ models are visualized in, respectively, light orange and light green in \autoref{fig:epe_roi_xcat}. The model in light orange was considered for the in-vivo experiments described in \autoref{sec:realtimeinvivo_new}, and shows acceptable performance for normal breathing and amplitude drifts, but is outperformed by the rank-2 models in the other two extreme scenarios. This indicates that more degrees of freedom would be favorable to model extreme changes in breathing pattern. Interestingly, the results on the rank-2 models show that a model trained on normal breathing can cleanly separate chest and abdominal motion-field components in the online phase.

The contribution of the offline reconstruction phase can be assessed by comparing reconstructions with the offline-reconstructed and ground-truth ${\bm{\Phi}}$ per breathing pattern. For rank-1 reconstruction this difference is minimal, but for rank-2 reconstructions the steep drop from the errors in the reconstructions with an offline reconstructed $\Phi$ to the reconstructions with ground-truth $\Phi$ shows that a large portion of the remaining errors can be attributed to the offline phase. An error below 0.75 mm is obtained in all scenarios with $R=2$ real-time MR-MOTUS reconstructions and a ground-truth $\bm{\Phi}$ (dark green). The dark green bars also show that the contribution of the rank-2 online reconstruction to the overall error is very minimal. In conclusion, the in-silico results indicate that higher-rank models with improved offline reconstruction quality could improve the overall quality of the proposed real-time reconstruction pipeline.
    
The spatial distribution of the EPEs in \autoref{fig:epe_spatial_distribution} shows acceptable errors within the ROI, the lungs and the liver. Higher errors are visible at the organ interfaces, which could be attributed to the disability of the smooth spline motion model to represent inter-organ discontinuities.

\subsection{In-vivo real-time 3D MR-MOTUS reconstructions}
\subsubsection{\color{nblue}Offline preparation phase}
\label{section:offlinemrmotusresults}
\begin{figure*}[tbp]
    \centering
    \includegraphics[width=.88\textwidth,height=.4\textwidth]{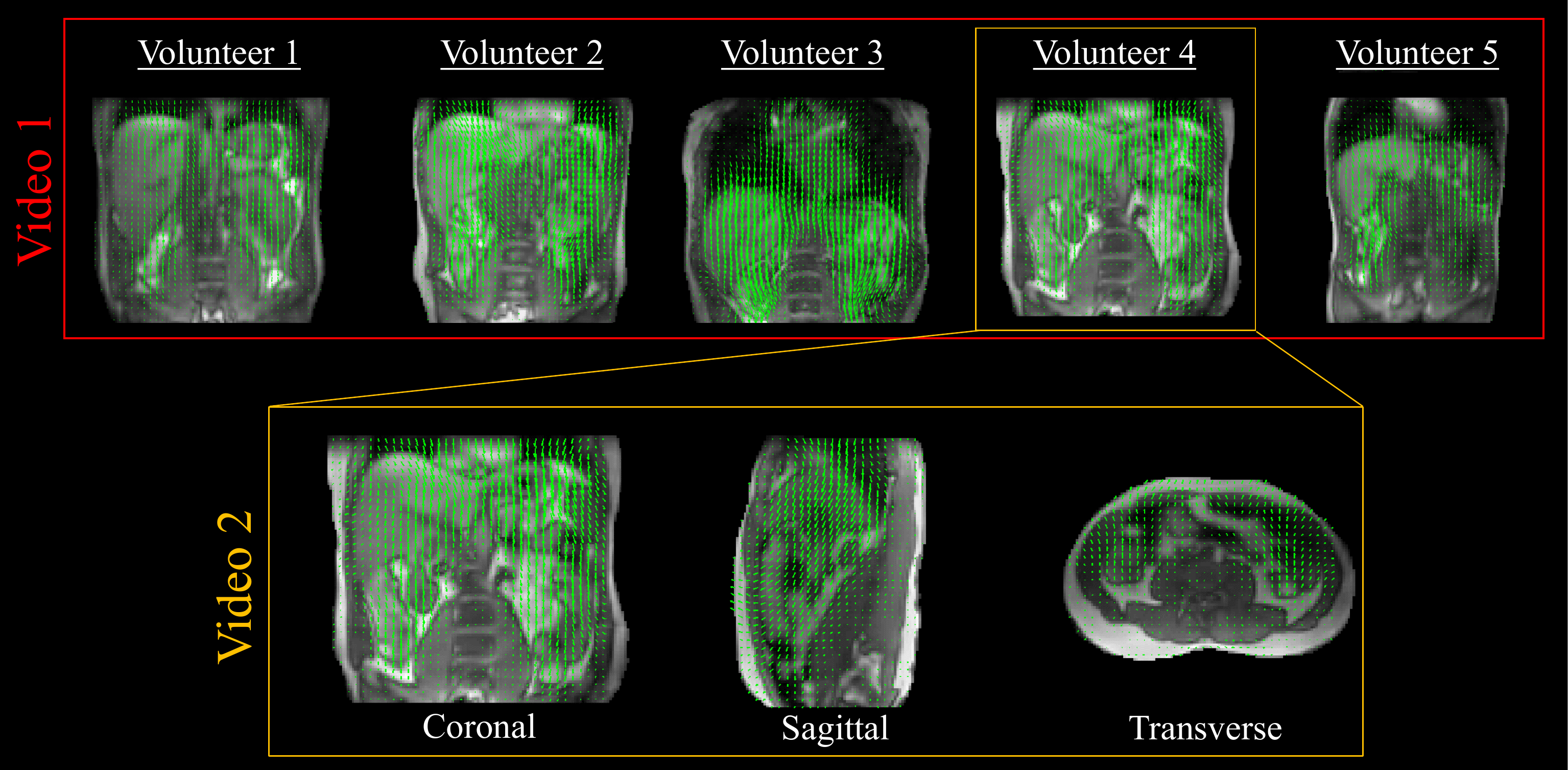}
    \caption{Snapshot of the reconstructed respiratory-resolved motion-fields, as described in \autoref{paragraph:offlinemrmotusmethods} and \autoref{section:offlinemrmotusresults}. These results should be viewed in the videos in the supporting files (see \hyperref[section:appendix]{Appendix}).}
    \label{fig:offline_mrmotus}
\end{figure*}

In the offline preparation phase, \autoref{eq:MRMOTUSoffline} was solved with respiratory-sorted data. \autoref{fig:offline_mrmotus} shows a snapshot of the reconstructed motion-fields for all volunteers in a coronal plane, and for volunteer 4 in three mutually orthogonal planes. These results should be viewed in the corresponding videos of the respiratory-resolved reconstructions: \hyperref[section:video1]{\texttt{Video1.mp4}} and \hyperref[section:video1]{\texttt{Video2.mp4}}. Please see the \hyperref[section:appendix]{Appendix} for an overview of all Videos. 
Little to no movement was reconstructed in organs not subject to respiratory motion such as the spine. Clear differences between the motion-fields for different volunteers can also be appreciated, most notably the large differences in the breathing motion amplitudes between e.g. volunteer 3 (large amplitude) and volunteer 5 (small amplitude).

\subsubsection{Online inference phase}
\label{section:invivorealtimeresults}
The real-time reconstructed motion-fields are visualized for volunteer 3 and volunteer 5, in respectively \hyperref[section:video3]{\texttt{Video3.mp4}} and \hyperref[section:video3]{\texttt{Video4.mp4}}. Overall realistically looking motion-fields are reconstructed. It can be observed that the motion-fields are mostly smooth in time, except for small high frequency fluctuations which are mostly visible in end-exhale for both volunteers. These fluctuations may be caused by hardware imperfections\cite{bruijnen2020prospective} or by sensitivity to cardiac motion, which manifest themselves as high frequency fluctuations on top of respiratory motion.

Supporting Information Figure S1 shows the dependency of the total processing time on the number of spokes per dynamic, and the latency target of 200 ms per dynamic for real-time MRgRT. Taking into account the fluctuations in the processing times, 14 spokes were selected to stay well below the latency target, while maximizing the number of spokes per dynamic. With 67 ms of acquisition time for 14 spokes, and 103 ms of reconstruction time per dynamic, this resulted in a 3D motion-field every 170 ms.

\subsection{In-vivo anatomical plausibility test with Jacobian determinants}
\label{section:validationjacdetresults}
\begin{figure*}[tbp]
    \centering
     \includegraphics[width=.88\textwidth,height=0.3\textwidth]{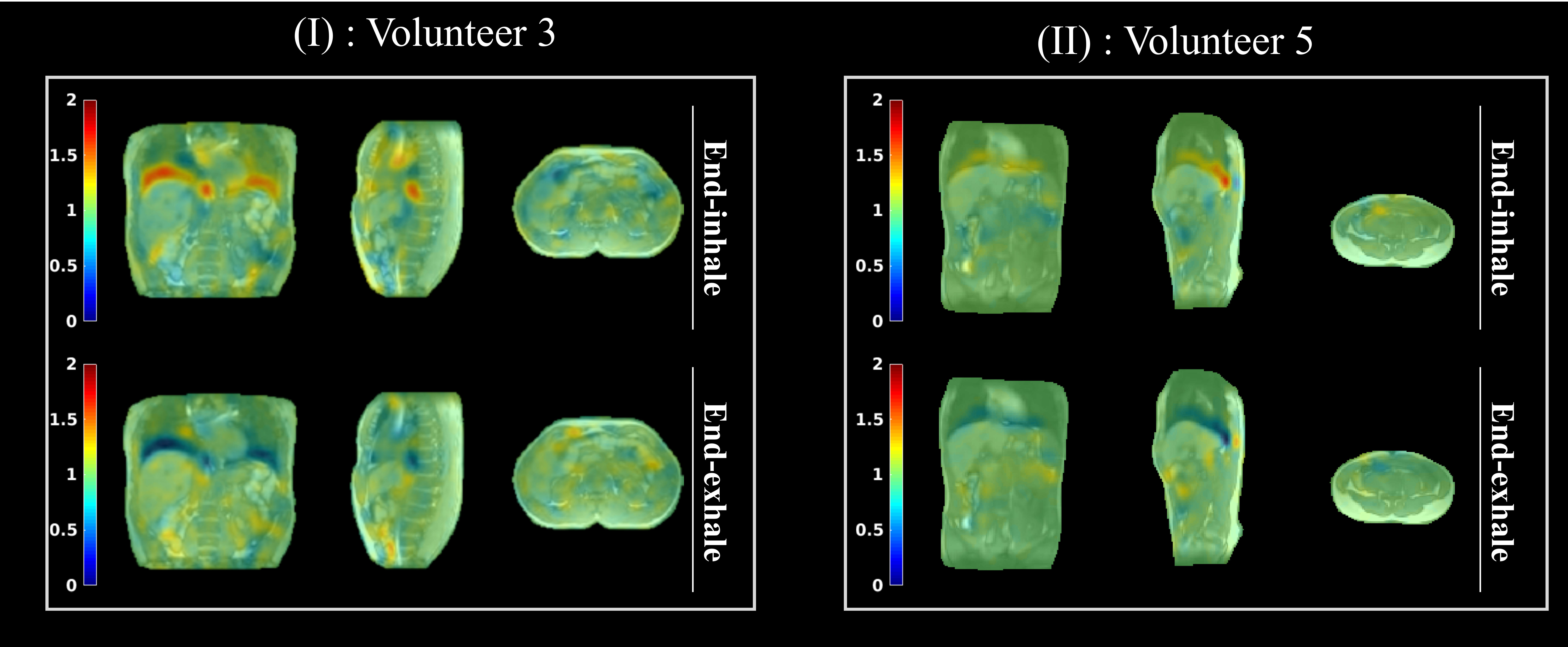}
    \caption{Validation of the motion-fields by means of their Jacobian determinants, as mentioned in \autoref{section:jacdet_validation_methods} and  \autoref{section:validationjacdetresults}. The value of the Jacobian determinant can be interpreted as the volume fraction after deformation with respect to a reference image. In this case the reference image is reconstructed in mid-position. Compression and expansion can be observed in the lungs, while organs such as the liver mostly preserve volume.}
    \label{fig:jacobiandet_validation}
\end{figure*}

\autoref{fig:jacobiandet_validation} shows the validation of the anatomical plausibility of the reconstructed motion-fields with its Jacobian determinant. For both volunteers it can be observed that most organs such as the liver preserve volume during deformation, which is in accordance with literature \cite{zachiu2018anatomically}. Bright red spots in the lungs indicate compression during inhalation, and bright blue spots indicate expansion during exhalation. Expansion and compression values are twice as high for volunteer 3 as for volunteer 5, indicating a relatively large breathing amplitude. Large values are also present at the interface between the top of the liver and the spine, where sliding motion occurs.

\subsection{In-vivo global accuracy test at high temporal resolution}
\label{section:validation_fh_results}
\begin{figure}[tbp]
    \centering
    \includegraphics[width=0.48\textwidth,height=.75\textwidth]{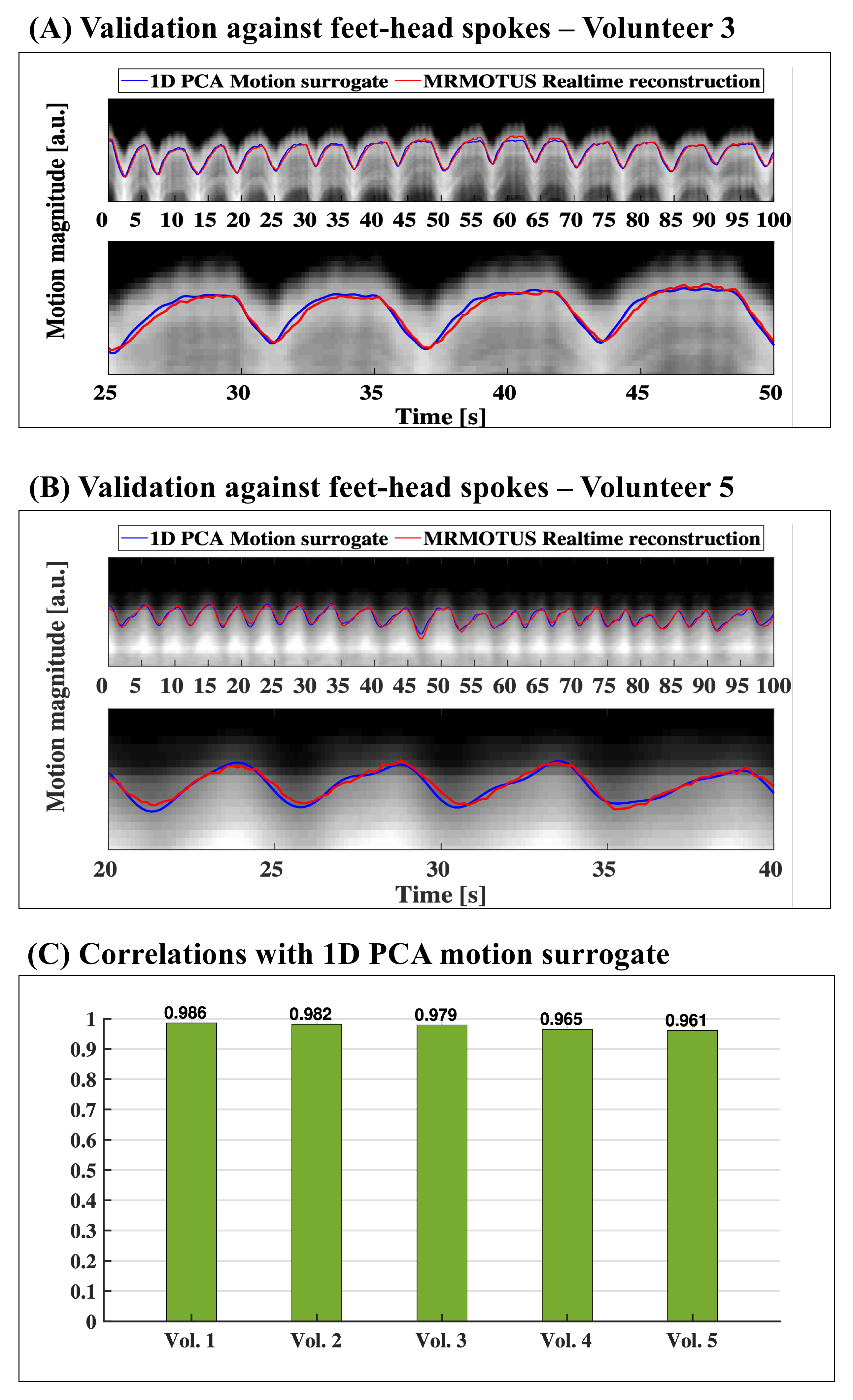}
    \caption{Results of the qualitative validation (A-B) and quantitative validation (C) of the reconstructed motion-fields at 6.7 Hz, as described in \autoref{section:fh_validation_methods} and \autoref{section:validation_fh_results}. (A-B) show clear visual similarity between the projection profiles, motion surrogate and MR-MOTUS reconstructions, and (C) substantiates this with high Pearson correlations of $0.975\pm 0.0110$. Scatter plots corresponding to the Pearson correlations can be found in Supporting Information Figure S2, and additional Bland-Altman analyses in Supporting Figure S3.
    }
    \label{fig:validation_fh}
\end{figure}

The self-navigation spokes along feet-head, that are interleaved with the golden-mean 3D radial kooshball acquisition, yield 1D respiratory motion information at high temporal resolution as projections of the whole excited FOV onto the feet-head axis. \autoref{fig:validation_fh}A and \autoref{fig:validation_fh}B show the validation of the reconstructed motion-fields with the projected profiles of the self-navigation spokes at 6.7 Hz and the surrogate signal extracted from these projections. Similar dynamic behavior of the projections, surrogate signal, and MR-MOTUS reconstructions can visually be observed for both volunteers, albeit that small high frequency oscillations remain present in the real-time MR-MOTUS reconstructions. Furthermore, \autoref{fig:validation_fh}B shows that the irregular breathing pattern of volunteer 5 is also reconstructed with MR-MOTUS. The dynamic behavior of the real-time reconstructions is quantitatively analyzed in \autoref{fig:validation_fh}C and Supporting Information Figure S2 by means of the Pearson correlation between the dynamic component $\bm{\Psi}$ of the real-time MR-MOTUS reconstructions and the 1D PCA motion surrogates. A linear correlation of $0.975 \pm 0.0110$ was found across all volunteers, further substantiating highly similar dynamic behavior between the 1D PCA motion surrogate and the real-time MR-MOTUS reconstructions. Additionally, Bland-Altman analyses are shown in Supporting Information Figure S3.

\subsection{Comparison with respiratory-resolved compressed sensing}
\begin{figure*}[tbp]
    \centering
    \includegraphics[width=0.8\textwidth,trim={1cm 0.5cm 1.2cm 0.2cm},clip]{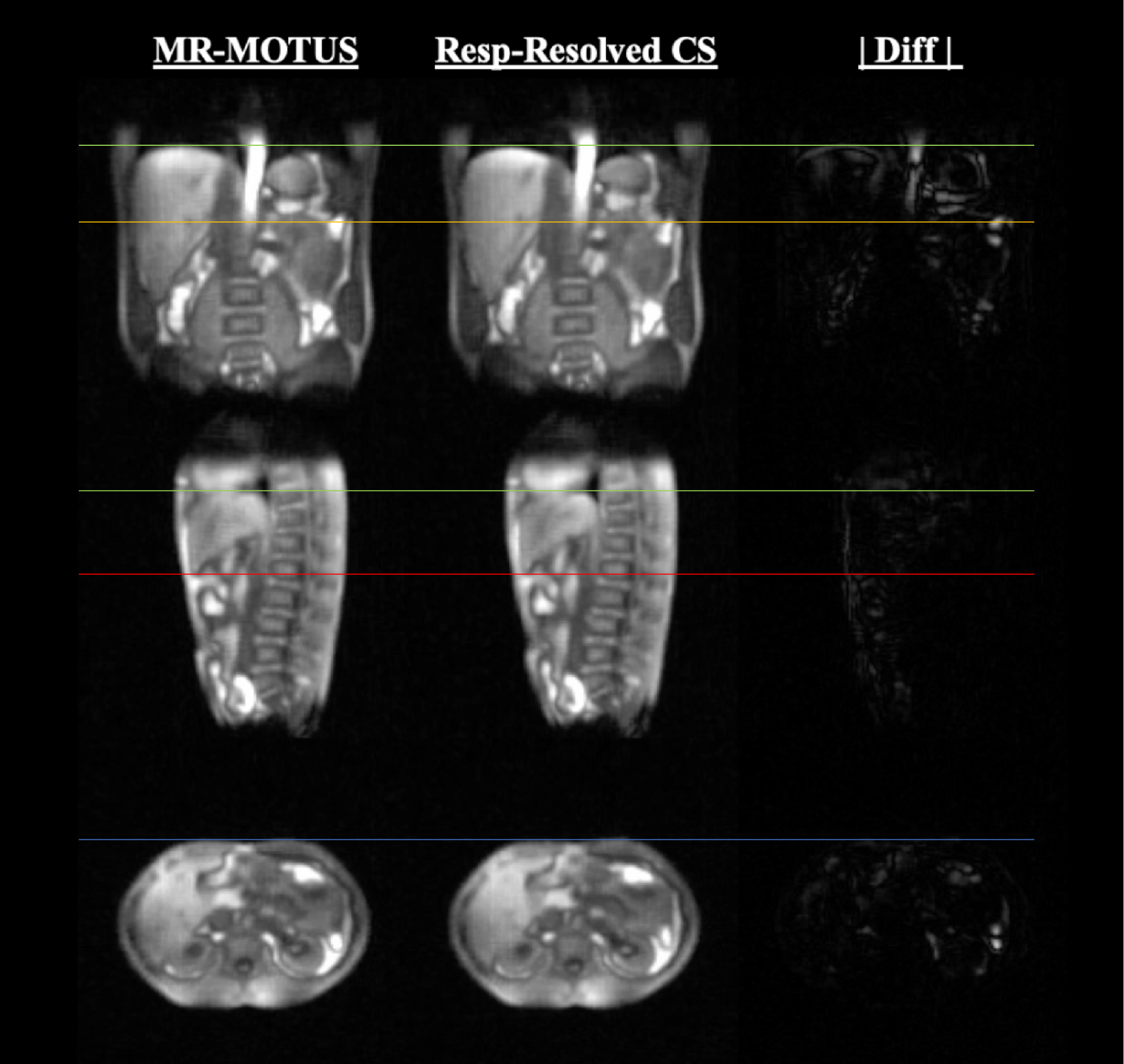}
    \caption{Snapshot of \hyperref[section:video5]{\texttt{Video5.mp4}}, showing a comparison between respiratory-resolved MR-MOTUS and compressed sensing reconstructions for volunteer 1. The end-inhale phase was visualized, which showed the largest differences between the two reconstructions. The colored horizontal lines compare the vertical positions of anatomical landmarks: top of the liver dome (green), liver vessel (orange), bottom of the liver (red) and the anterior side of the lower abdomen (blue).}
    \label{fig:video5_snapshot}
\end{figure*}

The comparison of the respiratory-resolved offline MR-MOTUS and compressed sensing (CS) reconstructions is shown in \hyperref[section:video5]{\texttt{Video5.mp4}} and \hyperref[section:video5]{\texttt{Video6.mp4}} for respectively volunteer 1 and volunteer 4. A snapshot of Video 5 in end-inhale is shown in \autoref{fig:video5_snapshot}, this respiratory phase showed most differences between the two reconstruction. The colored horizontal lines show minimal differences between locations of anatomical landmarks. From the videos it can be observed that the dynamics in the two reconstructions are very similar. From the last column it can be observed that only minimal differences remain present between the top of the liver. It can also be observed that some motion is reconstructed in the top of the spine, which is not visible in the CS reconstructions. The rest of the spine, however, remains static. Finally, the CS reconstructions contain pulsations in the aorta, which by construction cannot be visualized in the MR-MOTUS reconstruction. In general it should be noted that contrast variations between the two reconstructions will also contribute to the difference images in the last column.

\section{Discussion}

In light of MR-guided radiotherapy, we have presented a method to perform real-time deformable 3D respiratory motion estimation with 170 ms latency including both data acquisition and reconstruction. The proposed method relies on splitting the large-scale motion reconstruction problem formulated in \cite{Huttinga2021} into two parts: (I) a medium-scale offline preparation phase and (II) a small-scale online inference phase which exploits the results of the offline phase for real-time computations. The method was validated on free-breathing data of 5 volunteers, acquired with an Elekta Unity MR-Linac. 

The results show that motion-fields reconstructed from data acquired with an MR-Linac are anatomically plausible \cite{zachiu2018anatomically}, have high correlation with a 1D motion surrogate \cite{feng2016xd,Pang2014}, and can be reconstructed with a latency of 170 ms that is sufficient for real-time adaptive MR-guided radiotherapy \cite{keall2006management,Murphy2002}. Hence, the proposed framework could be of significant value in a future clinical workflow for adaptive real-time MR-guided radiotherapy. Moreover, the motion-fields reconstructed in real-time during radiotherapy treatments could also be used for retrospective radiation dose accumulation calculations\cite{Kontaxis2020}. Furthermore, the current framework relies on a golden-mean 3D radial trajectory acquisition such that data required for the offline MR-MOTUS preparations, and the pre-treatment 3D+t respiratory-resolved MRI required for the radiotherapy workflow \cite{Winkel2019}, could be acquired simultaneously.

The low-rank representation of respiratory motion \cite{Huttinga2021} is an important component of this work, which has shown to represent realistic respiratory motion-fields with few parameters. The apparent strength of this representation is in agreement with results reported in several other works in the context of motion estimation \cite{cai20153d,stemkens2016image,zhang2007patient,king2012thoracic,mishra2014initial,li2011pca,Borman2019,paganelli2018mri}. The strategy to split the motion-fields into spatial low-rank motion-field components and temporal motion-field components for fast inference was proposed before, e.g. for real-time CT-based motion estimation \cite{Li2010,Li2011}, and for MR-based motion estimation \cite{stemkens2016image,Borman2019}. However, a notable difference is that in this work and \cite{Huttinga2021}, the low-rank structure of motion-fields is enforced a-priori; the low-rank components are obtained with a model-based reconstruction that ensures consistency with 3D $k$-space data of a training phase. In the other works \cite{Li2010,Li2011,stemkens2016image,Borman2019}, the low-rank components are retrospectively obtained with PCA {\it after} a model-based reconstruction, and thus do not necessarily ensure consistency with data of the training phase. Similar approaches that reconstruct all low-rank components directly from the data were also proposed for dynamic MR-image reconstruction \cite{Liang2007,ong2019extreme}, but were not extended to real-time reconstructions.

There are two large differences between the proposed method and other methods in the context of real-time motion estimation: the dimensionality of the input data (1D, 2D, 2.5D or 3D) and the processing domain (image domain or $k$-space domain). The former can be reduced to increase the temporal resolution of 3D reconstructions. For example, several works use multiple orthogonal 2D-cine planes, i.e. 2.5D+t, to reconstruct volumes (motion-fields or images), achieving a frame rate of about 2 Hz excluding reconstruction \cite{stemkens2016image,Paganelli2018,Borman2019,Bjerre2013,tryggestad20134d}. Further reducing the dimensionality, \cite{brix2014three} achieved 5 Hz with 2D input data, and \cite{Feng2020} generated 3D volumes at 3.3 Hz (acquisition + reconstruction) with 1D input data. Slightly different type of methods are based on surrogate signal models \cite{Mcclelland2017,mcclelland2013respiratory,andreychenko2017thermal,Low2005,Tran2020} that - similarly to this work - also use a bi-linear motion model, but directly incorporate 1D surrogate signals in this model to infer 3D motion-fields from 1D input data. These methods can thus achieve high temporal resolution, but rely heavily on the quality of the motion surrogate. In contrast with the other works, MR-MOTUS uses 3D input data and thereby has the ability to take higher dimensional motion information into account, but possibly at a lower spatial resolution. The other large difference is the processing domain; MR-MOTUS fits motion-fields directly in the $k$-space domain, whereas most other methods fit in the image domain. Fitting directly in $k$-space has the advantage of being more flexible in terms of temporal resolution and dimensionality of input data, but fitting on single-channel $k$-space data comes at the cost of a reduction in SNR and an increased sensitivity to hardware imperfections that cause temporal signal fluctuations such as eddy-currents. We have empirically observed that both can be controlled well with the proposed regularization techniques. However, better control of e.g. eddy-currents on an MR-Linac can further improve the data quality \cite{bruijnen2018gradient}, and may thereby improve the results in this work. 

There are several other points that should be discussed. Firstly, the offline reconstruction times are currently in the order of minutes, of which most time is taken up by type-3 non-uniform FFTs (NUFFTs) \cite{huttinga2020mr}. Reconstruction times could thus be improved significantly by faster NUFFT computations. 

The data acquisition in the offline phase was not optimized; the 10 minutes used for the in-vivo results presented in this work was chosen on forehand to ensure sufficient data to reconstruct a reference image. This should be considered as a very conservative scenario to demonstrate the potential of the method. Results on the digital motion phantom and preliminary in-vivo results indicate that around 3 minutes of free-breathing data would be sufficient to perform the same reconstructions. In practice, the data acquisition may be performed simultaneously with the pre-treatment MRI that is required for radiotherapy \cite{Winkel2019}, or in the idle time during the treatment plan optimization, which takes 3-4 minutes in our current MR-Linac workflow. In the latter scenario no additional time would be added to the treatment. Finally, radiolucent coils with more SNR that are currently being developed \cite{Zijlema2019} could further improve our acquisition protocol. The optimization of the data acquisition protocol will be considered in a future work. 

There is also still room for improvement in the data acquisition in the online phase. For example, 14 spokes per dynamic and only 8 samples per spoke were sufficient to perform the real-time reconstructions, but more samples were acquired per spoke. Although the radial readouts considered in this work yielded good performance, the proposed framework is in theory not limited to a specific type of readout but in practice a readout is required that contains sufficient motion information. Thus, different, more efficient trajectories could further reduce the latency of the online phase. However, designing such a trajectory is not trivial; a trajectory that only traverses the 8 samples required for the online reconstructions across multiple spokes would be affected by different system imperfections (zeroth and first order eddy currents in particular) compared to the golden mean spokes. This could induce a discrepancy between the offline and online data that could potentially affect the real-time motion estimation. 

In this work a spatial motion-field basis was reconstructed in the offline phase, and was used in the online phase for fast inference. The spatial basis was reconstructed from respiratory-sorted data acquired over several minutes, and should therefore be able to represent a wide range of respiratory motion. The proposed framework, including both offline and online reconstruction phases, is in theory compatible with any pre-specified rank in the motion model that could be required to model a wide range of motion. The in-silico experiments show that a higher-rank model built for normal breathing can be used for real-time reconstructions of different breathing patterns, and in-vivo results in \autoref{fig:validation_fh}B also show good reconstructions for the heavily varying breathing pattern of volunteer 5. The current work used a rank-1 motion model for in-vivo reconstructions, since we empirically observed that larger motion models did not significantly improve in-vivo reconstruction quality, but did increase real-time computation times. However, to accurately model in-vivo respiratory motion on larger time scales, it may be required to either update the spatial basis during the real-time reconstructions or use higher-rank motion models as was done for the in-silico experiments. The in-silico results also indicate that improving the quality of the offline reconstructions of the less-dominant motion modes can most significantly improve the overall quality of the proposed real-time reconstruction pipeline. Future work will focus on improving reconstructions of higher-rank in-vivo motion models.

The regularization added to the real-time reconstructions was shown to reduce high-frequency oscillations, but small oscillations still remain present. We expect these are caused by either high-frequency physiological motion such as cardiac motion, or by hardware-related system imperfections such as eddy-currents that are known to affect the data quality of the Elekta Unity MR-Linac \cite{bruijnen2018gradient}. Furthermore, the real-time reconstructions were shown to have only 170 ms latency, but in a practical application even this delay will have to be compensated. Possible directions could be Gaussian Processes \cite{rasmussen2003gaussian} or other Bayesian filters \cite{chen2003bayesian}, to simultaneously perform adaptive filtering and short-term predictions.

In this work we have chosen to perform an extensive in-silico validation of the local reconstruction errors to get an impression of the potential weaknesses of the proposed framework. The in-vivo validation of 3D+t motion-fields at high temporal resolution, however, remains challenging, and we have chosen to evaluate the correlation with a 1D feet-head motion surrogate that can be extracted at high temporal resolution with PCA. This surrogate has been shown to be highly correlated with physical translation in feet-head direction \cite{Pang2014}. Results show high correlation between the global scaling of MR-MOTUS motion-fields and the PCA-based motion surrogate, indicating that the global dynamic behavior of the in-vivo MR-MOTUS motion-fields is very similar to a widely used motion surrogate. However, the PCA-based surrogate is one-dimensional, and therefore does not allow for a local 3D validation of the accuracy of the motion-fields in-vivo. To completely assess the in-vivo performance of the proposed method, more validations are required with e.g. with a higher dimensional image navigator, fiducial marker tracking, or external respiratory motion sensors \cite{Hoogeman2009}. 

A different possible direction for future work is the extension of the proposed framework to other types of motion. The flexibility of low-rank MR-MOTUS was demonstrated in \cite{Huttinga2021} by reconstructing both respiratory motion and head motion with the same low-rank motion decomposition model, but with a different number of components. The present work extends low-rank MR-MOTUS to real-time reconstructions without making direct assumptions on the type of motion, so the proposed real-time framework may be extended as was done in \cite{Huttinga2021}.

\section{Conclusion}
We have demonstrated real-time low-rank MR-MOTUS, a framework that can reconstruct 3D nonrigid motion-fields in real-time with a total latency of 170 ms. The proposed method was validated in-silico and evaluated on a hybrid 1.5T MR-linac, and could reconstruct motion-fields that are anatomically plausible and are highly correlated with respiratory surrogate signals. We anticipate that low-rank MR-MOTUS could provide a novel practical solution for real-time MR-guided abdominal radiotherapy in the future.

\appendices

\newpage
\label{section:appendix}


\section{Descriptions of supporting files}
All supporting files are available at \url{https://surfdrive.surf.nl/files/index.php/s/vz2xmwliglRmcjo}.
\subsection*{Video1.mp4, Video2.mp4}
\label{section:video1}
\noindent Offline reconstructed respiratory-resolved 3D+t motion-fields in a coronal plane for all volunteers (Video 1) and in all planes for volunteer 5 (Video 2). The moving images are obtained by warping the reference image with the motion-field. The visualized motion-fields (green arrows) show the displacement magnitude and direction projected on the selected planes.

\subsection*{Video3.mp4, Video4.mp4}
\label{section:video3}
\noindent Online time-resolved 3D+t motion-fields for volunteer 3 (Video 3) and volunteer 4 (Video 4), reconstructed at 16.2 Hz in 170 ms per dynamic, and visualized at 8.1 Hz. The moving images are obtained by warping the reference image with the motion-field. The visualized motion-fields (green arrows) show the displacement magnitude and direction projected on the selected planes.

\subsection*{Video5.mp4, Video6.mp4}
\label{section:video5}
\noindent Comparisons between offline-reconstructed respiratory-resolved MR-MOTUS and compressed sensing reconstructions, both resulting in respiratory-resolved volumetric images. The videos show results for volunteer 1 (Video 5) and volunteer 4 (Video 6). To obtain the MR-MOTUS images, the reference image was warped with the offline-reconstructed motion-fields.

\subsection*{SupportingFigures.pdf}
\noindent Supporting document with additional figures.
\label{section:si}

\section{Real-time implementation details}
\subsection{Preliminaries}
Assume a rank $R$ motion model, i.e. 
$$\bv{X}(\bv{r},t) = \bv{X}(\bv{r},0) + \bv{D}(\bv{r},t) = \bv{X}(\bv{r},0) +  \bm{\Phi}(\bv{r}) \bm{\Psi}^T(t),$$ for $\bv{X}(\bv{r},t),\bv{X}(\bv{r},0),\bv{D}(\bv{r},t)\in\mathbb{R}^{d \times 1}$, $\bm{\Phi}(\bv{r})=[\bm{\Phi}_{1}(\bv{r}),\dots,\bm{\Phi}_{R}(\bv{r})]\in \mathbb{R}^{d\times R}$, and $\bm{\Psi}(t)\in\mathbb{R}^{1 \times R}$. For the online inference phase, we will only deal with data and unknowns at a specific point in time, and for ease of notation we will therefore drop the continuous time-dependency and write $\bm{\Psi}(t)$ as $\bm{\Psi}_{t}:=[ \bm{\Psi}_{t1} , \dots, \bm{\Psi}_{tR} ] \in \mathbb{R}^{1 \times R}$ in what follows. Substitution of the motion model in the signal model \eqref{eq:forwardmodel} yields
\begin{alignat*}{2}
F(\bm{\Phi,\Psi}_t)[\bv{k}] &= \int_\Omega q(\bv{r})e^{-i 2 \pi \bv{k}^T\left(\bv{r} + \bm{\Phi}(\bv{r})\bm{\Psi}_t^T\right)} \infd \bv{r}, \quad \bv{k}\in\mathcal{K}_t,
\end{alignat*}
where $\mathcal{K}_t$ denotes the set of $k$-space coordinates assigned to the acquisition at time $t$. We have split the arguments in $F$ for ease of notation in the derivation of the derivatives in the next subsection. \subsection{Jacobian and Hessian of the forward model}
In the online inference phase we are interested in real-time reconstruction of $\bm{\Psi}_t$, given $\bm{\Phi}$ and data, which requires extremely fast evaluations of both forward model and derivatives with respect to the unknowns $\bm{\Psi}_t$. In this work we employ a Gauss-Newton algorithm, which requires a Jacobian matrix $\bv{J}$ with first order derivatives and approximates the Hessian matrix as $\bv{J}^H\bv{J}$. Alternatively, a Newton method can be followed by using the true Hessian matrix, but we have experienced minor improvement, and increased reconstruction times with the Newton method as opposed to the Gauss-Newton method. For clarity we derive both the Jacobian and true Hessian matrix. We first derive the required derivatives:
\begin{multline*}
\left(\frac{\partial^{a+b}{F}}{\partial  \left[\bm{\Psi}_{tm}\right]^a \partial \left[\bm{\Psi}_{tn}\right]^b }\right)[\bv{k}] = \int_\Omega q(\bv{r}) e^{-i 2 \pi \bv{k}^T \left(\bv{r} + \bm{\Phi}(\bv{r})\bm{\Psi}_t^T\right)}\cdot \\ \left[-i 2 \pi \bv{k}^T \bm{\Phi}_{m}(\bv{r})\right]^a \left[-i 2 \pi \bv{k}^T \bm{\Phi}_{n}(\bv{r})\right]^b \infd \bv{r},
\end{multline*}
where $\bv{k}_l \in \mathbb{C}^{d \times 1}$ is the $l$-th $d$-dimensional $k$-space coordinate in the sequence $\mathcal{K}_t$ of $k$-space coordinates in the dynamic at time $t$. Similarly as in the manuscript, we denote the total number of $k$-space coordinates at dynamic $t$ as $N_k:=|\mathcal{K}_t|$. Hence, the Jacobian $\bv{J}\in\mathbb{C}^{N_k \times R}$ and the Hessian matrix $\underline{\bv{H}}\in\mathbb{C}^{N_k \times R \times R} $ of $\bv{F}(\bm{\Phi},\cdot): \mathbb{R}^R \mapsto \mathbb{C}^{N_k}$, evaluated at $\bm{\Psi}_t$, are respectively computed as 
$$[\bv{J}]_{l,m} := \int_\Omega q(\bv{r}) e^{-i 2 \pi \left(\bv{k}_l\right)^T \left(\bv{r} + \bm{\Phi}(\bv{r})\bm{\Psi}_t^T\right)}\cdot \left[-i 2 \pi \left(\bv{k}_l\right)^T \bm{\Phi}_{m}(\bv{r})\right],$$
\begin{multline*}
[\underline{\bv{H}}]_{l,m,n} := \int_\Omega q(\bv{r}) e^{-i 2 \pi \left(\bv{k}_l\right)^T \left(\bv{r} + \bm{\Phi}(\bv{r})\bm{\Psi}_t^T\right)}\cdot \\ \left[-i 2 \pi \left(\bv{k}_l\right)^T \bm{\Phi}_{m}(\bv{r})\right] \left[-i 2 \pi \left(\bv{k}_l\right)^T \bm{\Phi}_{n}(\bv{r})\right] \infd \bv{r}.
\end{multline*}
Note that the Hessian of the vector-valued function $\bv{F}(\bm{\Phi},\cdot)$ is a 3-tensor. In what follows the notations of all tensors will be denoted by capital, underlined bold letters and in general we follow the tensor notations in \cite{cichocki2009nonnegative}.
\subsection{Jacobian and Hessian of the objective function}
We now assume subscripts denote tensor indices, and a colon selects all elements along a dimension, $\odot$ denotes element-wise multiplication, and $\overline{\times}_n$ $n$-mode tensor-vector products (i.e. dot-products along the $n$-th tensor dimension). Furthermore, we define $\bv{K}\in\mathbb{R}^{N_k \times d \times N_t}$ as the tensor with $N_k$ $d$-dimensional $k$-space coordinates at times $t\in[1,\dots N_t]$, and $\hat{\bm{\Phi}} \in \mathbb{R}^{N\times d \times R}$ as the discretization of $\bm{\Phi}(\bv{r})$. Finally, we define the following matrices that are required for efficient computations in the real-time reconstructions: 
\begin{alignat*}{2}
[\underline{\bv{L}}]_{:,:,m}&:= -i 2 \pi \left( [ \underline{\bv{K}}]_{:,:,t}\right) [\bm{ \hat{\Phi} }]_{:,:,m}^T,
\\
[\bv{G}]_{:,j}&:=-i 2 \pi \left([\underline{\bv{K}}]_{:,:,t}\right)\bv{r}_j, 
\\
\bv{A}_t&:=\text{exp}(\bv{G} + \underline{\bv{L}} \ \overline{\times}_3 \ \bm{\Psi}_t ),
\end{alignat*}
with $\bv{\underline{L}} \in \mathbb{C}^{N_k \times N \times R}, \bv{G} \in \mathbb{R}^{N_k \times N},\bv{A}_t\in\mathbb{C}^{N_k \times N},$ and denote the vectorization of $q(\bv{r})$ as $\bv{q}_0\in\mathbb{C}^{N\times1}$, with $N$ spatial points and $d$ spatial dimensions. Then, the evaluations of the forward model, Jacobian and Hessian at $\bm{\Psi}_{t}$ can respectively be written in tensor-vector form as
\begin{alignat}{2}\label{eq:discretehessianjacobians1}
\bv{F}(\cdot, {\bm{\Psi}_{t}}) &= \bv{A}_t \bv{q}_0 \quad  \in \mathbb{C}^{N_k \times 1} \\ \label{eq:discretehessianjacobians2}
[\bv{J}]_{:,m} &= \left(\bv{A}_t \odot [\underline{\bv{L}}]_{:,:,m} \right)\bv{q}_0 \quad  \in \mathbb{C}^{N_k \times 1} \\  \label{eq:discretehessianjacobians3}
[\underline{\bv{H}}]_{:,m,n} &= \left(\bv{A}_t \odot [\underline{\bv{L}}]_{:,:,m} \odot [\underline{\bv{L}}]_{:,:,n}\right) \bv{q}_0 \quad \in \mathbb{C}^{N_k \times 1 \times 1}.
\end{alignat}
The objective function for dynamic $t$ can now be formulated as 
$$E(\bm{\Psi}_t):= \lVert \bv{e}_t \rVert_2^2, \quad \bv{e}_t := \bv{A}_t\bv{q}_0 - \bv{s}_t \ \in \mathbb{C}^{N_k \times 1}. $$ The gradient $\bm{\nabla}_E$, true Hessian $ \bv{H}_E$, and Gauss-Newton Hessian approximation $ \hat{\bv{H}}_E$ of $E$ can be derived using \eqref{eq:discretehessianjacobians1}-\eqref{eq:discretehessianjacobians3} as \begin{alignat}{2}
 \bm{\nabla}_E &= 2\Re\left(\bv{J}^H\bv{r}_t \right) \ \in \mathbb{R}^{R \times 1} \\
\bv{H}_E &= 2\Re\left(\bv{J}^H\bv{J} + \underline{\overline{\bv{{H}}}} \ \overline{\times}_1 \ \bv{e}_t \right) \ \in \mathbb{R}^{R \times R} \\
\hat{\bv{H}}_E &= 2\Re\left(\bv{J}^H\bv{J}\right) \ \in \mathbb{R}^{R \times R},
\end{alignat}
where the superscript $H$ denotes the conjugate transpose. As argued in the manuscript, $R$ will typically be small, so these Jacobian and Hessian are very small, allowing for fast (sub-millisecond) processing.

\setcounter{algorithm}{-1}
\begin{algorithm}[h!]
\begin{algorithmic}[0]
\caption{Real-time MR-MOTUS Gauss-Newton algorithm.}
\STATE
\STATE \% Initialize solution variables 
\STATE $\bm{\Psi}_{t={M_\textrm{train}}}:=\bv{0} \hfill \% \ \bm{\Psi}_{t={M_\textrm{train}}}\in\mathbb{R}^{R \times 1}$
\STATE $ \bm{\hat{\Phi}} := \textrm{reshape}(\bm{\Phi},N,d,R) \hfill \% \ \bm{\hat{\Phi}}\in\mathbb{R}^{N \times d \times R}$
\STATE
\FOR{$t>M_\textrm{train}$}
\STATE
\STATE \% Pre-compute $t$-dependent tensors:
\STATE $[\underline{\bv{L}}]_{:,:,m}:= -i 2 \pi \left( [ \underline{\bv{K}}]_{:,:,t}\right) [\bm{ \hat{\Phi} }]_{:,:,m}^T \hfill \% \ \underline{\bv{L}} \in \mathbb{C}^{N_k \times N \times R}$
\STATE $[\bv{G}]_{:,j}:=-i 2 \pi \left([\underline{\bv{K}}]_{:,:,t}\right)\bv{r}_j \hfill \% \ \bv{G} \in \mathbb{C}^{N_k \times N}$
\STATE
\STATE \% Initialize Gauss-Newton iterations with solution at previous time-index:
\STATE $\bv{x}_t^{(1)} = \bm{\Psi}_{t-1} \hfill \% \ \bv{x}_t^{(1)} \in \mathbb{R}^{R \times 1}$
\STATE
\FOR{Gauss-Newton iterations $k=1\dots N_{GN}$}
\STATE
\STATE \% Compute the forward model matrix and the residuals at the current iterate
\STATE $\bv{A}_t:=\exp{\left( \bv{G} + \underline{\bv{L}} \ \overline{\times}_3 \ \bv{x}_t^{(k)}\right)} \hfill \% \ \bv{A}_t \in \mathbb{C}^{N_k \times N}$ 
\STATE $\bv{e}_t := \bv{A}_t \bv{q}_0 - \bv{s}_t \hfill \%  \ \bv{e}_t \in \mathbb{C}^{N_k \times 1}$ 
\STATE
\STATE \% Compute Jacobian of the forward model, and the gradient and approximated Hessian matrix of objective function w.r.t. solution variables
\STATE $[\bv{J}]_{:,m} := \left(\bv{A}_t \odot \underline{\bv{L}}_{:,:,m} \right)\bv{q}_0 \hfill \%  \ \bv{J} \in \mathbb{C}^{N_k \times R}$ 
 \STATE $\bm{\nabla}_E := 2\Re\left(\bv{J}^H\bv{e}_t \right) \hfill \%  \ \bm{\nabla}_E \in \mathbb{R}^{R \times 1}$ 
\STATE $\hat{\bv{H}}_E := 2\Re\left(\bv{J}^H\bv{J}\right) \hfill \%  \ \hat{\bv{H}}_E \in \mathbb{R}^{R \times R}$ 
\STATE
\STATE \% Compute Gauss-Newton step-direction
\STATE $\bm{\delta}_{\bv{x}}^{(k)} =  \textrm{argmin}_{\bm{\delta}} \ \lVert \hat{\bv{H}}_E \bm{\delta} + \bm{\nabla}_E\rVert_2^2 \hfill \% \ \bm{\delta}_{\bv{x}}^{(k)} \in \mathbb{R}^{R \times 1}$
\STATE
\STATE \% Update solution variable
\STATE $\bv{x}_{t}^{(k+1)}=\bv{x}_t^{(k)}+\bm{\delta}_{\bv{x}}^{(k)} \hfill \% \ \bv{x}_{t}^{(k+1)} \in \mathbb{R}^{R \times 1}$
\STATE
\ENDFOR
\STATE
\STATE \% Assemble 3D motion-field at time $t$
\STATE $\bm{\Psi}_t = \bv{x}_t^{(N)} \hfill \% \ \bm{\Psi}_t \in \mathbb{R}^{R \times 1} $
\STATE $\bv{D}_t = \bm{\Phi} \bm{\Psi}_t^T \hfill \% \ \bv{D}_t \in \mathbb{R}^{Nd \times R}$
\STATE
\ENDFOR
\end{algorithmic}
\caption{Gauss-Newton algorithm for real-time MR-MOTUS}
\label{algo:gnrealtimemrmotus}
\end{algorithm}

\bibliographystyle{latex/IEEEtran}

\end{document}